\documentclass[pdflatex,sn-mathphys-ay]{sn-jnl}
\usepackage{graphicx}
\usepackage{subcaption}				 
\graphicspath{{./Figs/}}				

\usepackage{amsmath,amssymb,amsfonts}
\usepackage{amsthm}
\usepackage{mathrsfs}

\usepackage{xcolor}
\usepackage{textcomp}				

\usepackage{multirow}
\usepackage{booktabs}				 

\usepackage[normalem]{ulem}		

\usepackage[title]{appendix}

\newcommand{\defeq}{\stackrel{\rm def}{=}}
\newcommand{\mB}{\mathcal{B}}
\newcommand{\mC}{\mathcal{C}}
\newcommand{\mD}{\mathcal{D}}
\newcommand{\mS}{\mathcal{S}}
\newcommand{\rthr}{r_{\text{thr}}}
\newcommand{\gamk}{{\gamma_k}}

\DeclareMathOperator*{\argmin}{arg\,min} 	

\title[Article Title]{Hedging via Perpetual Derivatives: Trinomial Option Pricing and
						Implied Parameter Surface Analysis}

\author*[ ]{Jagdish Gnawali} \email{jgnawali@ttu.edu}
\author[ ]{W. Brent Lindquist} \email{brent.lindquist@ttu.edu}
\author[ ]{Svetlozar T. Rachev} \email{zari.rachev@ttu.edu}
\affil[ ]{Department of Mathematics \& Statistics, Texas Tech University,
			1108 Memorial Circle, Lubbock, TX 79409}

\begin{document}
\maketitle
\noindent
{\bf Abstract.}
We introduce a fairly general, recombining trinomial tree model in the natural world.
Market-completeness is ensured by considering a market consisting of two risky assets,
a riskless asset, and a European option.
The two risky assets consist of a stock and a perpetual derivative of that stock.
The option has the stock and its derivative as its underlying.
Using a replicating portfolio, we develop prices for European options and generate
the unique relationships between the risk-neutral and real-world parameters of the model.
We discuss calibration of the model to empirical data in the cases in which the risky asset returns
are treated as either arithmetic or logarithmic.
From historical price and call option data for select large cap stocks,
we develop implied parameter surfaces for the real-world parameters in the model.



\section{Introduction}
Despite known limitations --
log-normal prices driven by Brownian motion;
absence of the drift term of the underlying in its option price;
assumption of the abilities to borrow any monetary amount at the risk-free rate
and trade assets of any monetary amount continuously in time with no transaction costs;
-- the Black-Scholes-Merton (BSM) model \citep{black_1973, merton_1973}
continues to serve as a fundamental reference tool in option pricing.
Our interest here is in discrete tree models,
which address option pricing without dealing with the machinery of stochastic integration theory.
As real trading occurs over (perhaps  very short, but nonetheless) discrete time intervals,
such models  avoid continuous-time assumptions and engender a more realistic pricing model.

As introduced by \cite{sharpe_1978} and formalized by \cite{cox_1979},
the basic discrete model, employing a recombining binomial tree,
was specifically designed to converge to the BSM model as $\Delta t \downarrow 0$.
Binomial pricing models have undergone continued development, including
providing faster rates of convergence \citep{leisen_1996} and
efficient computation of the ``Greeks'' \citep{tian_1993}, as well as
addressing the inclusion of stochastic volatility \citep{hilliard_1996, bates_1996},
skewness and kurtosis \citep{rubinstein_1998}, and
jump processes \citep{boyle_1986, bates_1996}.
\cite{kim_2019} extended the basic Cox-Ross-Rubenstein model to a new version with
time-dependent parameters.
\cite{hu_2020} further extended the \cite{kim_2019} binomial option pricing model to allow for
variable-spaced time increments.
 
Trinomial trees for option pricing were introduced by \cite{boyle_1986}.
As with original formulations of binomial models \citep{cox_1979, jarrow_1983},
trinomial trees were developed specifically to converge to the BSM option price formula in the
continuous-time limit.
By adding a third option to the pricing tree (that of no price change over a discrete time interval),
trinomial tree models provide a richer state space and
the potential for an improved rate of convergence to the BSM solution
(compared to binomial models).\footnote{
	See, however, the results of \cite{Chan_2009} which show that the Tian third-order moment binomial
	tree model outperforms eight other trinomial tree models.
}
A number of trinomial (and, by natural extension, multinomial) tree models have been developed subsequently
\citep{boyle_1988, boyle_1989, madan_1989, kamrad_1991, boyle_1994, florescu_2008, deutsch_2009, yuen_2010, ma_2015, langat_2019,
kim_2019}.
Convergence rate studies of trinomial models have been examined theoretically and numerically
\citep{ahn_2007, ma_2015, josheski_2020, lilyana_2021}.

A fundamental problem with the published trinomial (and multinomial) trees is that they
are defined directly in the risk-neutral world.
The free parameters of the model - the directional price change factors and probabilities -
are fit to the risk-neutral BSM model.
Consequently connection to crucial natural world parameters
(the price drift and directional change probabilities) are lost.
This connection is lost because no hedging is performed (i.e., no replicating portfolio is developed).
This issue was first addressed in \cite{kim_2019} for the specific case of the Cox-Ross-Rubenstein model.
However, their trinomial model is not market-complete.
The purpose of this paper is to address the market-completeness issue in the context of a
fairly general trinomial model.
To ensure market completeness, we work within a market consisting of
two risky assets, a riskless asset and a European contingent claim (call or put option).
The market uncertainty is driven by a single Brownian motion.
To ensure this, the two risky assets consist of a stock and a derivative based on that stock.
As we wish to price the option for any possible maturity date,
the stock derivative is chosen to be a perpetual derivative.
We develop the replicating portfolio producing risk-neutral pricing.
The resulting unique relationship between the risk-neutral and real world parameters enables
computation of real world implied parameter values.

In Section \ref{sec:pd}, we briefly review the price dynamics of the stock perpetual derivative
\citep{shirvani_2020, Lindquist_2024}.
In Section \ref{sec:tri} we develop our general trinomial tree model,
establishing the unique relationship between risk-neutral and real-world parameters.
In Section \ref{sec:params} we discuss calibration of the model's natural-world parameters to
empirical data in the cases in which asset returns are either arithmetic or continuous (i.e., log-returns).
Application of the model to empirical data is presented in Section \ref{sec:emp}.
We consider historical stock and option prices for three large cap stocks and develop implied surfaces
for the following parameters: volatility, price drift, price change probabilities, and the risk-free rate.
Conclusions are presented in Section \ref{sec:Conclusion}.

\section{The Perpetual Derivative}\label{sec:pd}

Consider a market containing a stock $\mS$ having price dynamics
\begin{equation}\label{eq:St}
	dS_t = \mu_t S_t dt+\sigma_t S_t dW_t ,
\end{equation}
where $W_t$ is a standard Brownian motion, $\mu_t$ is a drift and $\sigma_t$ is a volatility. The dynamics of a riskless asset (bond) ${\mB}$ is
\begin{equation}\label{eq:Bt}
	dB_t = r_{f,t} B_t dt,
\end{equation}
where $r_{f,t}$ is a risk-free rate.
Let $\mD$ denote a perpetual derivative of $\mS$ whose price, $g(S_t,t)$,
is governed by the It{\^o} process
\begin{equation}\label{eq:ito}
	dg_t = \left( \frac{\partial g_t}{\partial t} + \mu_t S_t \frac{\partial g_t}{\partial S_t}
		+ \frac{\sigma^2_t S_t^2}{2}\frac{\partial^2 g_t}{\partial S_t^2} \right) dt
		+ \sigma_t S_t \frac{\partial g_t}{\partial S_t} dW_t ,
\end{equation}
ensuring that uncertainty in the prices of $\mS$ and $\mD$ are driven by the same Brownian motion.
To ensure that $\mD$ can be priced, form a replicating portfolio
$\pi_t^{(\mD)} = a_t S_t + b_t B_t - g_t$.
Requiring $\pi_t^{(\mD)} = 0$ and $d \pi_t^{(\mD)} = 0$
leads, in the standard way, to the BSM PDE for the price dynamics of $\mD$,
\begin{equation}\label{eq:gt}
	r_{f,t} g_t = \frac{\partial g_t}{\partial t} + r_{f,t} S_t \frac{\partial g_t}{\partial S_t}
		+ \frac{\sigma^2_t S_t^2}{2}\frac{\partial^2 g_t}{\partial S_t^2}  ,
\end{equation}
with initial data $g_0(S_0,0)$.
\cite{Lindquist_2024} investigated separable solutions to \eqref{eq:gt} and
showed the existence of a one-parameter family of solutions.
Of these solutions, the price process
\begin{equation}\label{eq:pd}
	g_t(S_t,t) = S_t^{-\delta_t}, \qquad \delta_t = \frac{2r_{f,t}}{\sigma^2_t},
\end{equation}
for the perpetual derivative $\mD$  has the dynamics
\begin{equation}\label{eq:dgpd}
	dg_t = \mu_\delta g_t dt + \sigma_\delta g_t d W_t, \qquad
	\mu_\delta =  (1 + \delta_t) r_{f,t} - \mu_t \delta_t, \quad \sigma_\delta = -\delta_t \sigma_t ,
\end{equation}
ensuring that the uncertainty in $S_t$ and $g_t$ are driven by the same Brownian motion.
This is also the form of the perpetual derivative (assuming the time-independent parameters)
	used by \cite{shirvani_2020}.\footnote{
	Let $\xi \in \mathbb{R}$ be a parameter and $\cal{V}^{\xi}$ denote a perpetual derivative having the
	price process $\boldsymbol{V}_t^{\xi} = S_t^{\xi}\beta_t^{\gamma}, t\geq 0$,
	where $\gamma=\frac{1-\xi}{r_f}\left[r_f+\frac{1}{2}\xi\sigma^2\right]$.
	Then the price process $\boldsymbol{V}_t^{\xi}$ discounted by a riskless bond rate
	is a martingale under the EMM $\mathbb{Q} \sim \mathbb{P}$
	and thus the security $\cal{V}^{\xi}$  can be traded within the BSM market model.
	The log-return of this perpetual derivative is a linear combination of the log-returns of the
	underlying stock $\cal{S}$ and the bond $\cal{B}$.
	When $\xi = -2r_f / \sigma^2$, then $\gamma = 0$ and the perpetual derivative price becomes
	independent of the bond price.
}

\section{Trinomial Tree Model} \label{sec:tri}
\begin{figure}[htp]
	\centering
	\includegraphics[width=\textwidth]{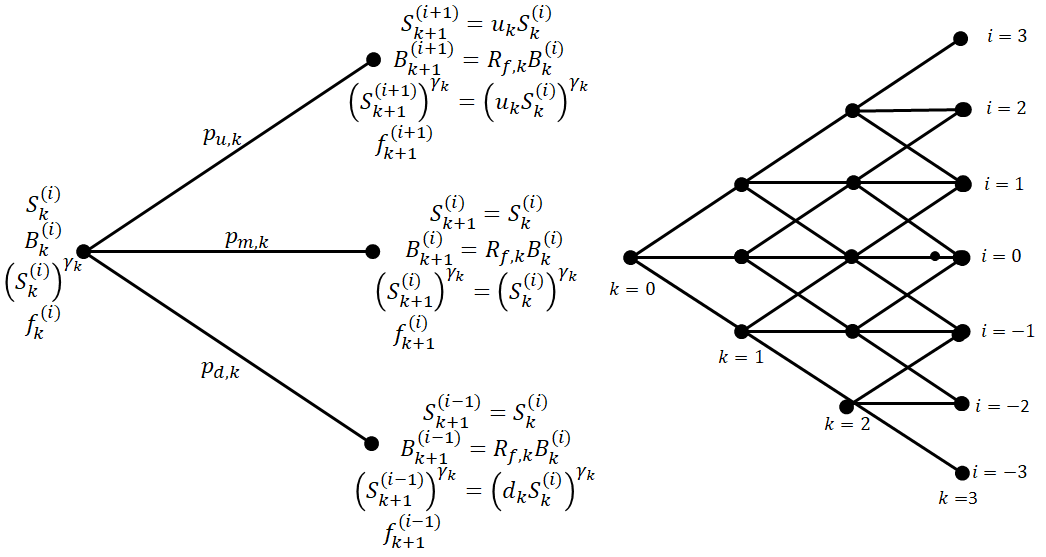}
	\caption{A trinomial tree showing (left) the pricing notation for the fundamental unit of the tree
	and (right) the time step $k$ and level $i$ indexing for a tree with three time steps.}
	\label{fig:tri_tree}
\end{figure}

Consider the market  $\{\mS, \mD, \mB, \mC\}$, where $\mC$ is a European contingent claim (option).
We model the price development of $\mS$, $\mD$ and $\mB$ on a trinomial tree
and use a replicating portfolio under no-arbitrage conditions to determine the price of $\mC$.
We develop a general trinomial pricing model first,
and then consider the two special cases in which returns are treated
either as arithmetic and logarithmic.
For simplicity we assume a constant time increment $\Delta t = T/N$ for the lattice,
where $T$ is the maturity date of the option.
The notation for the general trinomial pricing tree is summarized in Fig.~\ref{fig:tri_tree}.

On the ``fundamental unit'' of the trinomial tree,\footnote{
		This is often referred to as a ``single period'' tree.
		However, a single period tree would imply $k = 0$.
		We prefer the designation fundamental unit,
		as the tree is assembled by replication of this unit.
		In the binomial tree literature, it has become convention to adopt
		$S_{k+1}^{(u)}$ and	$S_{k+1}^{(d)}$	as the generic price changes.
		This convention does not extend naturally to multinomial trees.
		The level indexing  $S_{k+1}^{(i+1)}$, $S_{k+1}^{(i)}$  and  $S_{k+1}^{(i-1)}$
		employed here does extend naturally to multinomial (including binomial) trees, and we prefer it.
		To be consistent with a general multinomial tree nomenclature,
		our price change probabilities should be written $p_{u,k} \rightarrow p_{+1,k}$,
		$p_{m,k} \rightarrow p_{0,k}$, and $p_{d,k} \rightarrow p_{-1,k}$.
		We confess to being inconsistent in adopting the most general notation.
	}
the stock price follows the discrete process,
\begin{equation}\label{eq:Sp}
	S_{k+1}^{(i)} =
	\begin{cases}
	   \begin{aligned}
		S_{k+1}^{(i+1)}  &= S_k^{(i)} u_k &&\text{ w.p. } p_{u,k}, \\
		S_{k+1}^{(i)}      &= S_k^{(i)}     &&\text{ w.p. } p_{m,k}, \\
		S_{k+1}^{(i-1)}  &= S_k^{(i)} d_k &&\text{ w.p. } p_{d,k},
	   \end{aligned}
	\end{cases}
\end{equation}
where we adopt the shortened notation $S_k^{(i)} = S_{k\Delta t}^{(i)}$, $u_k = u_{k\Delta t}$,
$p_{u,k} = p_{u,k\Delta t}$, etc., $k = 0, 1, ...,N$.
The price change probabilities in the natural world are determined by independent,
trinomially distributed random variables
$\zeta_k$ satisfying: $p_{u,k} = P(\zeta_k = 1)$, $p_{m,k} = P(\zeta_k = 0)$,
and $p_{d,k} = P(\zeta_k = -1)$ where
$p_{u,k} + p_{m,k} + p_{d,k} = 1$, $k = 1, ..., N$.
The pricing trees in this trinomial model are adapted to the discrete filtration
\begin{equation}\label{eq:filt}
	\mathbb{F}^{(N)} = \left\{ {\cal F}^{(N,k)} = \sigma( \zeta_1, ..., \zeta_k ),
	\  k = 1, ..., N;\  {\cal F}^{(N,0)} = \{ \emptyset, \Omega\} \right\}.
\end{equation}
The probabilities $p_{u,k}$, $p_{m,k}$ and $p_{d,k}$ are $ {\cal F}^{(N,k)}$-measurable.
The perpetual derivative price follows the discrete process
\begin{equation}\label{eq:Dp}
	\left(S_{k+1}^{(i)}\right)^\gamk =
	\begin{cases}
	   \begin{aligned}
		\left(S_{k+1}^{(i+1)} \right)^\gamk &= \left(S_k^{(i)}\right)^\gamk u^\gamk_k &&\text{ w.p. } p_{u,k}, \\
		\left(S_{k+1}^{(i)}\right)^\gamk      &= \left(S_k^{(i)}\right)^\gamk		       &&\text{ w.p. } p_{m,k}, \\
		\left(S_{k+1}^{(i-1)} \right)^\gamk &= \left(S_k^{(i)}\right)^\gamk d^\gamk_k &&\text{ w.p. } p_{s,k},
	   \end{aligned}
	\end{cases}
\end{equation}
where, for notational simplicity, we denote $ \gamk = -\delta_k = -2r_{f,k}/\sigma^2_k$.
The dynamics of the bond price is
\begin{equation}\label{eq:Bp}
		B_{k+1}^{(i)} =
	\begin{cases}
	   \begin{aligned}
		B_{k+1}^{(i+1)}  &= B_k^{(i)} R_{f,k} &&\text{ w.p. } p_{u,k}, \\
		B_{k+1}^{(i)}      &= B_k^{(i)} R_{f,k} &&\text{ w.p. } p_{m,k}, \\
		B_{k+1}^{(i-1)}  &= B_k^{(i)} R_{f,k} &&\text{ w.p. } p_{d,k}.
	   \end{aligned}
	\end{cases}
\end{equation}

At time $t = k \Delta t$,
let $a_k$, $b_k$ and $c_k$ represent the number of respective shares of $\mS$, $\mB$ and $\mD$
held in a portfolio
used to replicate the price of the option ${\mC}$ having  ${\mS}$ and  ${\mD}$ as underlying.
Over the single time--step $k \rightarrow k+1$, the arbitrage--free, replicating portfolio obeys
\begin{subequations}
\begin{align}
	a_k  S_k^{(i)}    + b_k  B_k^{(i)}      + c_k \left(S_k^{(i)}\right)^\gamk 		 &= f_k^{(i)} ,  \label{eq:rp_0} \\
	a_k  S_k^{(i)} u_k + b_k B_k^{(i)} R_{f,k}  + c_k \left(S_k^{(i)}\right)^\gamk u^\gamk_k &= f_{k+1}^{(i+1)}, \label{eq:rp_up} \\
	a_k S_k^{(i)}    + b_kB_k^{(i)}    R_{f,k}  + c_k \left(S_k^{(i)}\right)^\gamk	 		 &= f_{k+1}^{(i)}, \label{eq:rp_md} \\
	a_k  S_k^{(i)} d_k + b_k B_k^{(i)} R_{f,k} + c_k \left(S_k^{(i)}\right)^\gamk d^\gamk_k &= f_{k+1}^{(i-1)}. \label{eq:rp_dn}
\end{align}
\end{subequations}
Solution of the system \eqref{eq:rp_up} - \eqref{eq:rp_dn}, determines the terms
$a_k  S_k^{(i)}$, $b_k  B_k^{(i)}$ and $c_k \left(S_k^{(i)}\right)^\gamk$.
From \eqref{eq:rp_0}, the recursive formula for the option price is then
\begin{equation} \label{eq:f0}
		f_k^{(i)} = R_{f,k}^{-1} (q_{u,k} f_{k+1}^{(i+1)} + q_{m,k} f_{k+1}^{(i)} + q_{d,k} f_{k+1}^{(i-1)} ),
\end{equation}
where the risk-neutral probabilities are
\begin{subequations}
\begin{align}
	q_{u,k}  &= \frac{(d^\gamk_k - d_k)(R_{f,k}-1)}{D_{1,k}}, \label{eq:qu}\\
	q_{m,k} &= 1 - q_{u,k} - q_{d,k} =\frac{u^\gamk_k (R_{f,k} - d_k) + R_{f,k} (d_k - u_k) + d^\gamk_k (u_k - R_{f,k})}{D_{1,k}}, \label{eq:qm}\\
	q_{d,k}  &=  \frac{(u_k - u_k^\gamk) (R_{f,k} - 1)}{D_{1,k}}, \label{eq:qd}\\
  \text{where} \quad 
	D_{1,k} & = (u_k - 1) d_k^\gamk - (u_k - d_k) + (1 - d_k) u_k^\gamk .\label{eq:D1}
\end{align}
\end{subequations}

It is straightforward to show that
$u_k q_{u,k} + q_{m,k} + d_k q_{d,k} = R_{f,k}$ and $u_k^\gamk q_{u,k} + q_{m,k} + d_k^\gamk q_{d,k} = R_{f,k}$.
Consequently
\begin{equation}\label{eq:pd}
	\pi_{u,k} \defeq \frac{q_{u,k}}{R_{f,k}}, \qquad \pi_{m,k} \defeq \frac{q_{m,k}}{R_{f,k}}, \qquad \pi_{d,k} \defeq \frac{q_{d,k}}{R_{f,k}}
\end{equation}
are the risk-neutral, single time step, price deflators:
\begin{align*}
	B_k^{(i)} &=  (\pi_{u,k}+ \pi_{m,k} + \pi_{d,k}) R_{f,k} B_k^{(i)} , \\
	S_k^{(i)} &= (\pi_{u,k} u_k + \pi_{m,k} +\pi_{d,k}  d_k) S_k^{(i)}, \\
	\left(S_k^{(i)}\right)^\gamk &= (\pi_{u,k} u_k^\gamk + \pi_{m,k} + \pi_{d,k} d_k^\gamk) \left(S_k^{(i)}\right)^\gamk.
\end{align*}

\section{Parameter Calibration and Continuous--Time Limits}\label{sec:params}

In order to calibrate the parameters to real data,
it is necessary to assume a form for the price change parameters
$u_k$, $d_k$ and $R_{f,k}$.
These forms must be self-consistent.
For arithmetic returns, the self-consistent modeling of the parameters is
\begin{equation}\label{eq:artn}
	u_k = 1+U_k,\qquad d_k = 1+D_k,  \qquad  R_{f,k} = 1 + r_{f,k} \Delta t,
\end{equation}
while for log-returns the parameters are modeled as
\begin{equation}\label{eq:lrtn}
	u_k = e^{U_k}, \qquad d_k = e^{D_k}, \qquad R_{f,k} = e^{r_{f,k}\Delta t}.
\end{equation}
In either case, the no-arbitrage condition requires $D_k < r_{f,k} \Delta t < U_k$.
From \eqref{eq:Sp}, the returns (whether arithmetic or logarithmic) are given by
\begin{equation}\label{eq:tri_rk}
	r_k = \left \{
	\begin{aligned} 
		U_k, &\text{ w.p. } p_{u,k} , \\
		0,      &\text{ w.p. } p_{m,k},\\
		D_k,&\text{ w.p. } p_{d,k}.
	\end{aligned}
	\right.
\end{equation}
We consider first the calibration of the natural world price change probabilities to historical data.
Let $\{ r_j, j = k-L+1, ..., k \}$ denote a historical record (i.e. a ``window of length L'') of return
(arithmetic or logarithmic) as appropriate data for $\cal{S}$.
Consider the threshold values $\rthr^+ \gtrsim 0$ and $\rthr^- \lesssim 0$.
Denote by: $L_{u,k}$ the number of these historical instances when $r_j \ge \rthr^+$;
$L_{m,k}$ the number when $\rthr^- < r_j < \rthr^+$; and
$L_{d,k}$ the number when $r_j \le \rthr^-$.
The natural probabilities can then be estimated from the historical data as
\begin{equation}\label{eq:pdmu}
p_{u,k} = \frac{L_{u,k}}{L}, \qquad p_{m,k} =  \frac{L_{m,k}}{L}, \qquad  p_{d,k}  = 1 - p_{u,k} - p_{m,k}.
 \end{equation}

The parameters $U_k$ and $D_k$ are estimated by setting the conditional first and second moments of $r_k$
to the instantaneous mean and variance of the historical return series,
\begin{subequations}
\begin{align}
	E \left[r_{k+1}|S_k^{(i)} \right] &= U_k p_{u,k} + D_k p_{d,k} =  \mu_k^{(r)} \Delta t , \label{eq:condE} \\
	\text{Var} \left[r_{k+1}|S_k^{(i)} \right] &= U_k^2 p_{u,k} + D_k^2 p_{d,k} - \left( E \left[r_{k+1}|S_k^{(i)} \right] \right )^2
		= \left(\sigma_k^{(r)}\right)^2  \Delta t. \label{eq:condV}
\end{align}
\end{subequations}
The instantaneous mean and variance are estimated using the same historical window
as for the price change probabilities.
Evaluating \eqref{eq:condE} and \eqref{eq:condV}  from \eqref{eq:tri_rk} produces
\begin{subequations}
\begin{align}
	U_k 
	&= \frac{1}{1 - p_{m,k}}
		\left\{ E \left[r_k|S_{k-1}^{(i)} \right]
			+ \sqrt{ \frac{ p_{d,k}  } { p_{u,k}  } } \sqrt{ \text{Var} \left[r_k|S_{k-1}^{(i)} \right] - p_{m,k} E \left[r_k^2|S_{k-1}^{(i)} \right] }
		\right\}   \nonumber \\
	&= \frac{1}{1 - p_{m,k}}
		\left\{ \mu_k^{(r)}\Delta t
			+ \sqrt{ \frac{ p_{d,k}  } { p_{u,k}  } }
				\sqrt{ (1-p_{m,k}) \left( \sigma_{k}^{(r)} \right)^2 \Delta t
					- p_{m,k} \left( \mu_k^{(r)} \Delta t \right)^2 }
		\right\} , \label{eq:U} \\
 	D_k 
 	&= \frac{1}{1 - p_{m,k}}
		\left\{ E \left[r_k|S_{k-1}^{(i)} \right]
			- \sqrt{ \frac{ p_{u,k}  } { p_{d,k}  } } \sqrt{ \text{Var} \left[r_k|S_{k-1}^{(i)} \right] - p_{m,k} E \left[r_k^2|S_{k-1}^{(i)} \right] }
		\right\} \nonumber \\
 	&= \frac{1}{1 - p_{m,k}}
		\left\{ \mu_k^{(r)}\Delta t
			- \sqrt{ \frac{ p_{u,k}  } { p_{d,k}  } }
				\sqrt{ (1-p_{m,k}) \left( \sigma_{k}^{(r)} \right)^2 \Delta t
					- p_{m,k} \left( \mu_k^{(r)} \Delta t \right)^2 }
		\right\} . \label{eq:D}
\end{align}
\end{subequations}
When $p_{m,k} = 0$, $p_{u,k} \equiv p_k$, $ p_{d,k} = 1 - p_k$, and \eqref{eq:U}, \eqref{eq:D}
reduce to the binomial tree solutions,
\begin{equation*}
 	U_k = \mu_k^{(r)} \Delta t + \left[ \frac{1-p_k}{p_k} \right]^{1/2} \sigma_k^{(r)} \sqrt{\Delta t}, \qquad
 	D_k = \mu_k^{(r)} \Delta t - \left[ \frac{p_k}{1-p_k} \right]^{1/2} \sigma_k^{(r)} \sqrt{\Delta t}.
\end{equation*}
The risk-neutral probabilities are computed from \eqref{eq:qu} to \eqref{eq:D1} using 
\eqref{eq:U}, \eqref{eq:D} and either
\eqref{eq:artn} or \eqref{eq:lrtn}, as appropriate.

From \eqref{eq:Sp}, for arithmetic returns, the conditional mean and the variance of the stock price are
\begin{align*}
    \mathbb{E} \left[S_{k+1}| S_k^{(i)}\right] &= S_k^{(i)}\left( 1 + E \left[r_{k+1}|S_k^{(i)} \right] \right)
   		 = S_k^{(i)}\left( 1 + \mu_k^{(r)} \Delta t \right), \\
  \mathbb{V} \left[S_{k+1}| S_k^{(i)}\right] &=\left( S_k^{(i)}\right)^2 \text{Var} \left[r_{k+1}|S_k^{(i)} \right]
  		= \left( S_k^{(i)}\right)^2 \left(\sigma_k^{(r)} \right)^2 \Delta t  ,
\end{align*}
and the instantaneous drift and variance of the risky asset  price and arithmetic return processes are identical,
\begin{equation}
	\mu_k =\mu_k^{(r)}, \qquad \sigma_k^2 = \left(\sigma_k^{(r)} \right)^2 .
\end{equation}
However, for log-returns, there is no simple relation between conditional mean and the variance
of the stock price
\begin{subequations}
\begin{align}
	\mathbb{E} \left[S_{k+1}| S_k^{(i)}\right] &= S_k^{(i)}\left( p_{u,k} e^{U_k} + p_{m,k} + p_{d,k} e^{D_k} \right) , \label{eq:lcondE}\\
	\mathbb{V} \left[S_{k+1}| S_k^{(i)}\right] &=\left( S_k^{(i)}\right)^2 \left(  p_{u,k} e^{2 U_k} + p_{m,k} + p_{d,k} e^{2 D_k}  \right)
  		- \left( E \left[S_{k+1}|S_k^{(i)} \right] \right )^2 , \label{eq:lcondV}
\end{align}
\end{subequations}
and the conditional mean and variance of the log-return \eqref{eq:condE}, \eqref{eq:condV}.
Under the assumption that terms of $o(\Delta t)$ can be neglected,
the exponentials in \eqref{eq:lcondE} and \eqref{eq:lcondV} can be expanded
producing the results
\begin{align*}
	\mathbb{E} \left[S_{k+1}| S_k^{(i)}\right] &= S_k^{(i)}\left\{ 1 + \left(  \mu_k^{(r)} + \frac{ \left( \sigma_k^{(r)} \right)^2 } { 2} \right) \Delta t \right\} , \\
	\mathbb{V} \left[S_{k+1}| S_k^{(i)}\right] &=\left( S_k^{(i)}\right)^2  \left( \sigma_k^{(r)} \right)^2 \Delta t .
\end{align*}
Thus to terms of $O(\Delta t)$, for log--returns,
the instantaneous drift and variance of the price of the risky asset $\cal{S}$ are
\begin{equation*}
\mu_{k\Delta t} =\mu_{k\Delta t}^{(r)}+\frac{(\sigma_{k\Delta t}^{(r)})^2}{2},
\qquad \sigma_{k\Delta t}^2=(\sigma_{k\Delta t}^{(r)})^2 .
\end{equation*}

We consider the continuous time $\Delta t \downarrow 0$ limits of the trinomial tree price processes.
Let $\lim_{\Delta t \downarrow 0} k \Delta t = t \in [0,T]$.
As $\Delta t \downarrow 0$,
$\mu_k^{(r)} \rightarrow \mu_t^{(r)}$, $\sigma_k^{(r)} \rightarrow \sigma_t^{(r)}$,
$r_{f,k}^{(r)} \rightarrow r_{f,t}$; and $\gamma_k \rightarrow \gamma_t$,
where we assume that the second derivatives of $\mu_t^{(r)}$, $\sigma_t^{(r)}$, $ r_{f,t}$,
and $\gamma_t$ are continuous on $[0,T]$.\footnote{
	We impose sufficient conditions.
}
A non-standard invariance principle \citep{Davydov_2008} can be used to show that,
under arithmetic returns,
the pricing tree \eqref{eq:Sp} generates a stochastic process which converges weakly
in $D[0,T]$ topology \citep{Skorokhod_2005}
to the cumulative return process $R_t$ determined by
$$
	dR_t = dS_t / S_t = \mu_t^{(r)} dt + \sigma_t^{(r)} dW_t.
$$
Under log-returns,
the pricing tree \eqref{eq:Sp} generates a c{\`a}dl{\`a}g process
which converges weakly in $D[0,T]$ topology to a continuous
diffusion process governed by the stochastic differential equation
$$
dS_t = \left( \mu_t^{(r)} + \frac{1}{2} \left( \sigma_t^{(r)} \right)^2 \right) S_t dt + \sigma_t^{(r)} S_t dW_t .
$$
In either case, the deterministic bond pricing tree \eqref{eq:Bp} converges uniformly to
$$
	B_t = B_0 e^{\int_0^t r_{f,s}\,ds }.
$$

\subsection{Estimation of $\rthr^-$ and $\rthr^+$}\label{sec:r_est}
Estimation of the threshold values $\rthr^-$ and $\rthr^+$ are critical for determining
the range of returns that
define $p_m$, i.e.  that indicate ``no (significant) change in the stock price''.
We estimate these thresholds using hypothesis testing on mean values, as follows.
Let $\{r_{t-L+1}, ..., r_t\}$ denote a window of historical returns.

Consider the value $p > 0$ basis points.
Let $\mathbb{S}_p = \{r_{t-k} \,|\, 0 \le r_{t-k} \le 10^{-4} p\}$
denote the sample of historical non-negative returns having value $\le 10^{-4} p$.
Let $\mu_p$ and $s_p$ denote, respectively, the mean and standard deviation of the sample $\mathbb{S}_p$.
Perform a $t$-test for the null hypothesis $H_0: \mu_p = 0$ versus the alternate $H_a: \mu_p > 0$.\footnote{
	Use of the $t$-test for sample means assumes that the center of the distribution of
	returns is well approximated by a normal distribution.
}
Given a fixed significance level $\alpha$, for  a sufficiently small value $\delta_p > 0$ of $p$,
$H_0$ will not be rejected.
If we examine a sequence of values $p_j = j\, \delta p$, $H_0: \mu_{p_j} = 0$
will not be rejected for $j = 1, ..., J^+$,
while $H_0: \mu_{p_{J^{+} +1}} = 0$ will be rejected.
We set the threshold $\rthr^+ = \mu_{p_{J^+}}$.\footnote{
	The procedure adopted here was motivated by the computation of VaR and CVaR values.
	The value $10^{-4} p_j$ is analogous to a VaR value,
	while $\mu_{p_j}$ is analogous to the related CVaR value.
	In this view, $\rthr^+ = \mu_{p_{J^+}}$ is the largest ``CVaR'' value for which the null hypothesis
	$H_0: \mu_{p_{J^+}} = 0$ is not rejected.
}

By considering a sequence of basis points $p_j = j \, \delta p$ with $\delta_p < 0$,
and defining the samples $\mathbb{S}_{p_j} = \{r_{t-k} \,|\, 10^{-4} p_j \le r_{t-k} \le 0\}$,
we can perform  $t$-tests for the null hypotheses $H_0: \mu_{p_j} = 0$
versus the respective alternates $H_a: \mu_{p_j} < 0$.
The first rejection of the $H_0: \mu_{p_j} = 0$ will occur at some value $j = J^- +1$.
We set the threshold $\rthr^- = \mu_{p_{J^-}}$.

Changing the significance level $\alpha$ will affect the values of $J^-$ and $J^+$.
As illustrated in Section~\ref{sec:emp}, we use a very stringent significance level.

\subsection{Estimation of Extreme Values for $\rthr^{-}$ and $\rthr^{+}$}\label{sec:r_ext}
As $p_d + p_m + p_u = 1$, there are only two independent price-change probabilities,
which can be expressed as either $\{\rthr^-, \, \rthr^+\}$ or $\{p_d, \, p_m\}$.\footnote{
	We focus on $p_d$ rather than $p_u$ as investors react more strongly to market downturns
	than to market upturns.
}
In Section~\ref{sec:emp} our focus will be on computing implied parameter surfaces by fitting
the computation of theoretical option prices to published option prices.
By holding $p_d$ constant (equivalently, $\rthr^-$), one can compute implied values for $p_m$
(equivalently, $\rthr^+ - \rthr^-$).
By holding $p_m$ constant, one can compute implied values for $p_d$.
Implied parameter values reflect (as a function of time to maturity and moneyness) the views of the market
towards the value of that parameter.
One of our investigations in Section~\ref{sec:emp} will be on the market view of the probability
of extreme downturns.
For this view, we will consider extreme returns below the conditional value at risk
$\text{CVaR}_{\beta}$\footnote{
	We ignore the convention that defines values of conditional value at risk corresponding
	to losses as positive.
}
and above the conditional value of return $\overline{\text{CVaR}}_{\beta}$.

Extreme price change probabilities can be computed from \eqref{eq:pdmu} by setting
$\rthr^- = \text{CVaR}_{\beta}$ and $\rthr^+ = \overline{\text{CVaR}}_{\beta}$.
We consider $\beta = 0.01$ corresponding to the first percentile.

\section{Application to Empirical Data}\label{sec:emp}

Using a window $\{t-L+1, ..., t\}$ of historical data, we utilize the trinomial tree model
with arithmetic returns
to compute call option price surfaces for day $t$ (Section \ref{sec:OP}).\footnote{
	We compute prices for European options.
	When we compute implied parameter values, we will use empirical prices for American options.
	As call option prices for European and American options are identical, but put option prices differ,
	we consider only call option prices and implied parameter values based on call options.
}
Using published option price data for $t$, we compute implied parameter surfaces, specifically for
volatility (Section \ref{sec:IV}), mean (Section \ref{sec:IM}),  risk-free rate (Section \ref{sec:IRf}),
and price change probabilities (Section \ref{sec:IP}).
Implied surfaces for $p_d$ and $p_m$ based on the extreme thresholds are provided in
Section~\ref{sec:IPext}.
The historical window covered the trading days from 01/16/2020 through  01/16/2024.
We considered three stocks, Apple (AAPL), Amazon (AMZN) and Microsoft (MSFT).\footnote{
	Stock and option price data obtained from Yahoo Finance. Accessed on 01/17/2024.
}
The risk-free rate $r_{f,t}$ for the date 01/16/2024 was taken from the US Treasury 10-year yield curve.
For each stock, the initial price used in computing options was the adjusted closing price on 01/16/2024.
To eliminate dividend artifacts, all returns were computed from adjusted closing prices.

As noted in Section~\ref{sec:r_est}, estimation of values for $\rthr^-$ and $\rthr^+$
depends on the significance level $\alpha$ used in hypothesis tests.
Using $\delta p = 1$ basis point, Table~\ref{tab:rthr} in the appendix shows how $\rthr^-$ and $\rthr^+$
 vary for these three stocks for values of $\alpha \in \{0.05, 0.01, 0.005, 0.001\}$.
 We use the values $\rthr^-$ and $\rthr^+$ obtained for $\alpha = 0.001$,
 which establishes a strong criterion for rejecting the null hypothesis.

\subsection{Option prices} \label{sec:OP}
Let $G^{(\text{emp})}(S_0,T_i,K_j)$, $i = 1, ..., I$, $j = 1, ..., J$,
denote published prices for a call option having $\mathcal{S}$ as the underlying.
Let
$G^{(\text{th})} \left( S_0,T_i,K_j;\sigma, \mu_t^{(r)} , r_{\text{thr}}, r_{f,t} \right)$
denote the respective theoretical option prices computed from the trinomial tree. 
We computed theoretical call option prices on $t = $ 01/16/2024
for maturity times corresponding to trading dates $t+T$, $T = 1, ..., T_I$ 
and strike prices $K \in \{K_1, ..., K_J\}$.
\begin{table}[htb]
	\caption{Parameter values computed from the historical returns}
	\label{tab:params}
	\centering
	\begin{tabular}{l ccc ccc cc l cc }
	\toprule
	Stock   & $S_0$	& $\mu$			   & $\sigma$& $p_d$ & $p_m$	& $p_u$	& $r_{f,t}$	& yearly	& daily \\
	\midrule
	AAPL   & 192.94	& $1.09 \cdot 10^{-3}$ & 0.0212	& 0.473	& 0.00995	& 0.517	& 3 Mo	& 0.0545	& $5.83 \cdot 10^{-4}$ \\
	AMZN & 153.16	& $7.69 \cdot 10^{-4}$ & 0.0238	& 0.477	& 0.00498	& 0.518	& 10 Yr	& 0.0407	& $1.09 \cdot 10^{-4}$\\
	MSFT  & 388.15	& $1.10 \cdot 10^{-3}$ & 0.0205	& 0.470	& 0.00796	& 0.522	& \omit	& \omit	& \omit\\
	\bottomrule
	\end{tabular}
\end{table}

\begin{figure}[htbp]
	\centering
	\includegraphics[width=0.32\linewidth]{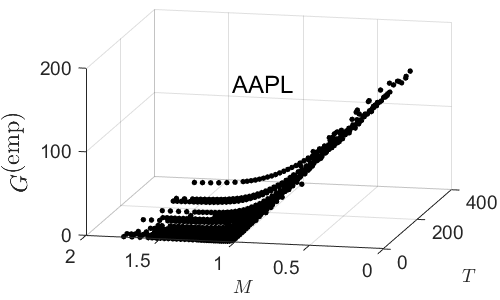}
	\includegraphics[width=0.32\linewidth]{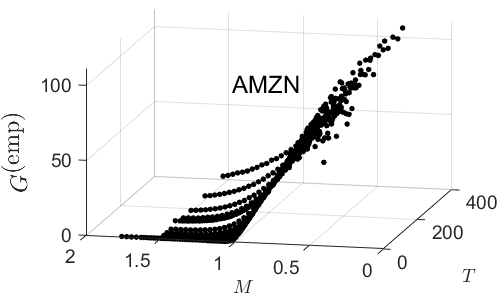}
	\includegraphics[width=0.32\linewidth]{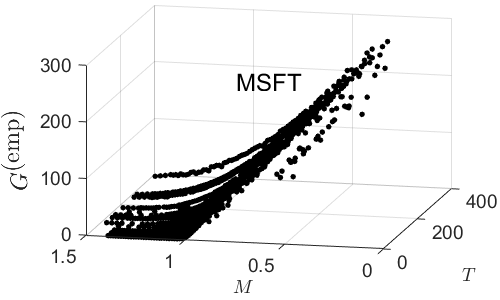}
	\includegraphics[width=0.32\linewidth]{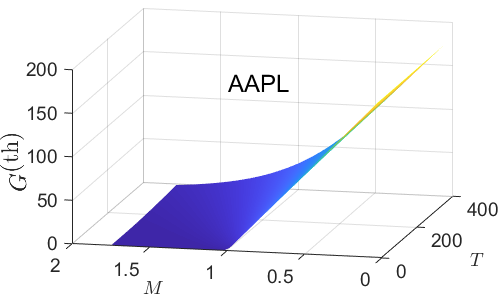}
	\includegraphics[width=0.32\linewidth]{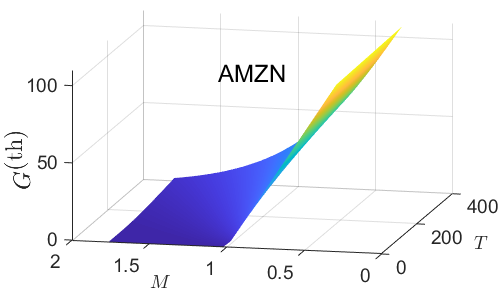}
	\includegraphics[width=0.32\linewidth]{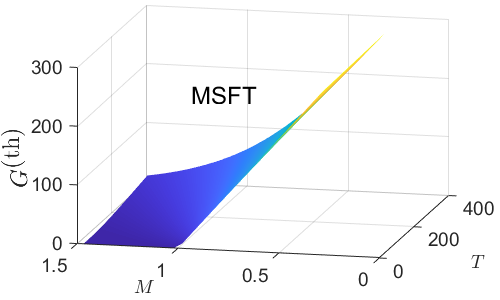}
	\includegraphics[width=0.32\linewidth]{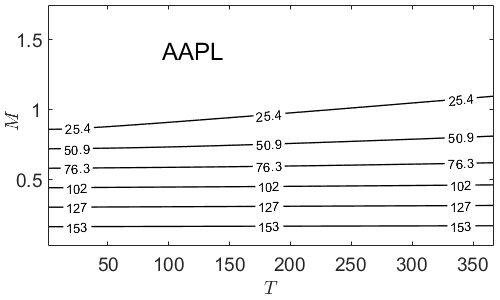}
	\includegraphics[width=0.32\linewidth]{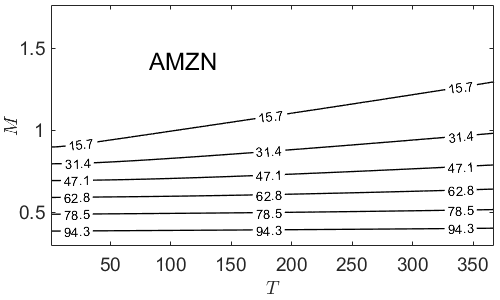}
	\includegraphics[width=0.32\linewidth]{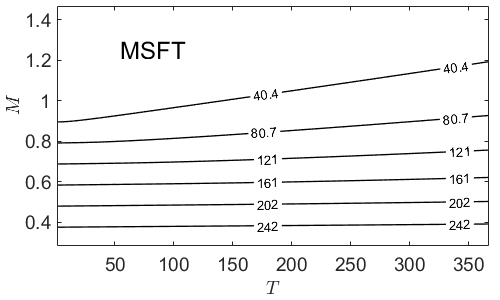}
	\caption{(top row) Empirical call option prices for the three assets.
		(middle row) Computed (theoretical) call option prices.
		(bottom row) Surface contours of the theoretical price surface projected on the ($T,M$) plane.
	}
	 \label{fig:CP}
\end{figure}
Option prices were plotted as functions of $T$ and moneyness values $M = K/S_0$,
with $S_0$ denoting the stock price on 01/16/2024.
The parameters  $p_{u,t}$, $p_{m,t}$, $p_{d,t}$, $\mu_{u,t}^{(r)}$,  and $\sigma_{u,t}^{(r)}$
were estimated from the historical returns over the period 01/16/2020 through 01/14/2024
as described in Section~\ref{sec:params}.
Table~\ref{tab:params} provides the values for $S_0$ and the estimated parameters.
The parameter values $p_d$, $p_m$ and $p_u$ are computed based on the $\alpha = 0.001$ threshold values
presented in Table~\ref{tab:rthr}.
Table~\ref{tab:params} also provides the US Treasury 10-year and 3-month yield rates for 01/16/2024.
The 3-month rate will be used in Section~\ref{sec:IRf}.

The empirical call option prices $G^{(\text{emp})}(S_0,T,K)$ are displayed as 3D scatterplots\footnote{
	Scatterplots are employed to show how sparsely in $T$ and $K$ the empirical data is populated.
}
in Fig.~\ref{fig:CP}.
The theoretical call option prices
$G^{(\text{th})} \left( S_0,T_i,K_j;\sigma, \mu_t^{(r)} , r_{\text{thr}}, r_{f,t} \right)$
based on the historical parameter values in Table~\ref{tab:params} are plotted as surfaces\footnote{
	The theoretical option price data can be computed to arbitrary fineness in values of $T$ and $M$.
}
in Fig.~\ref{fig:CP}.
Also plotted are contour levels of $G^{(\text{th})} $ projected on the $T,M$ plane.
For constant values of maturity $T$, we note the non-monotonicity of $G^{(\text{emp})}$ with $K$,
in contrast to the monotonicity of $G^{(\text{th})}$.

\subsection{Implied volatility} \label{sec:IV}

\begin{figure}[htbp]
	\centering
	\includegraphics[width=0.32\linewidth]{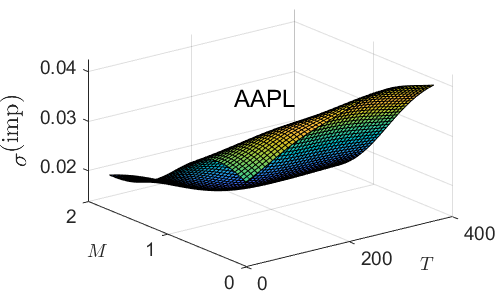}
	\includegraphics[width=0.32\linewidth]{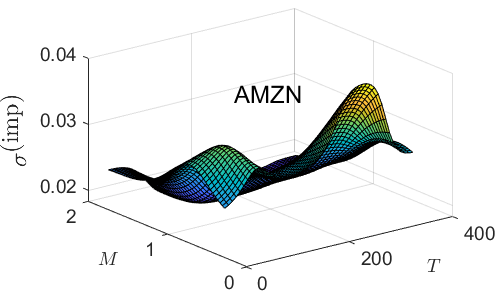}
	\includegraphics[width=0.32\linewidth]{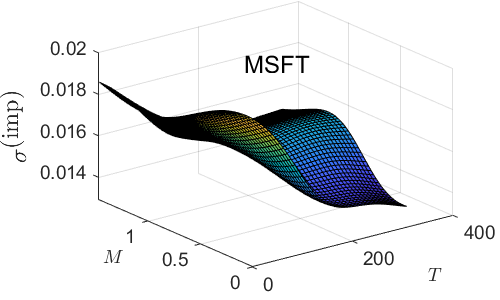}
	\includegraphics[width=0.32\linewidth]{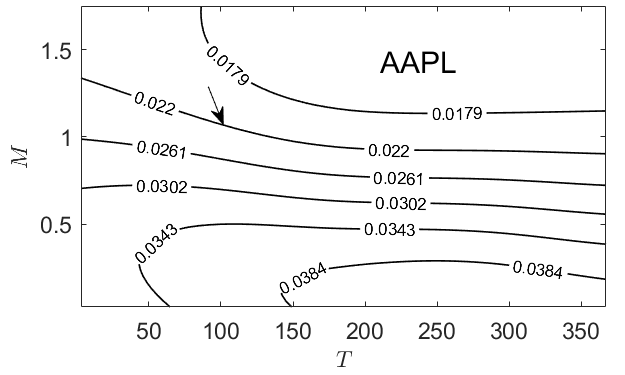}
	\includegraphics[width=0.32\linewidth]{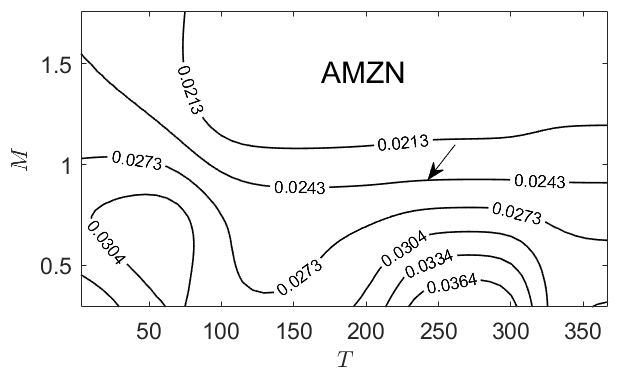}
	\includegraphics[width=0.32\linewidth]{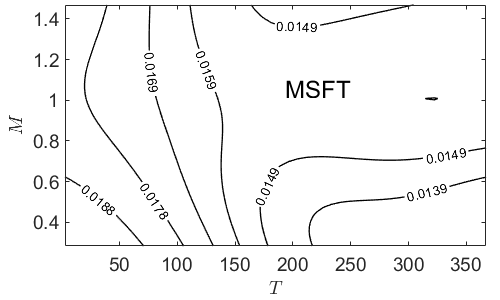}
	\caption{(top row) Computed implied volatility surfaces $\sigma^{(\text{imp})}(T,M)$
		for call options for the three assets.
		(bottom row) Surface contours projected on the ($T,M$) plane.
		Arrows indicate contour closest to historical volatility.
	}
	 \label{fig:IV}
\end{figure}

The implied volatility is given by
\begin{equation}\label{eq:IV}
\sigma^{(\text{imp})}(T_i,K_j) = \argmin_\sigma
	\left (
		\frac{G^{(\text{th})} \left( S_0,T_i,K_j;\sigma, \mu_t^{(r)} , r_{\text{thr}}, r_{f,t} \right) - G^{(\text{emp})}(S_0,T_i,K_j) }
			{G^{(\text{emp})}(S_0,T_i,K_j)}
	\right )^2 ,
\end{equation}

and is computed for all pairs of values $(T_i,K_j)$ for which there is empirical data.
Using a Gaussian kernel smoother, implied volatility values are then computed for all coordinates
$(T_i,K_j)$, $i = 1, ..., I$, $j = 1, ..., J$.\footnote{
	Without further mention,
	computation of implied values for all pairs of values $(T_i,K_j)$ for which there is empirical data,
	and the use of a Gaussian kernel smoother to ``fill in'' implied values for all possible $(T_i,K_j)$
	coordinate pairs will be performed for each implied parameter discussed below.
}
The resultant implied volatility surfaces for call options are shown in Fig.~\ref{fig:IV}.
Also plotted are contour levels of $\sigma^{(\text{imp})} $ projected on the $T,M$ plane.

Based on the contour plots, one can (approximately) position the contour associated with the
historical value of $\sigma$ in Table~\ref{tab:params}.
For AAPL, the closest contour plotted is 0.022.
For values of $M$ higher than one in the contour (i.e. further out-of-the-money),
the option values are predicated on lower volatility than the historical;
for smaller values of $M$ (further in-the-money), the views of option traders are based on
higher volatility.
In other words, the historical-valued contour separates more confident (out-of-the-money)
option price projections from less confident (into-the-money) option price projections.
For AMZN, the results are roughly the same, with a contour level of 0.0238
(approximated by the 0.0243 contour shown) providing the separation.
In contrast, for MSFT, all volatility contour levels lie below the historical value of $\sigma = 0.0205$.
For MSFT, option traders have greater confidence in the option price projections
over the projected range of $M$ and $T$, than the historical value of $\sigma$ would indicate.
Thus, (for the date 01/16/2024) option traders were projecting lower future volatility for MSFT
than the historical value.

\subsection{Implied mean} \label{sec:IM}
\begin{figure}[htbp]
	\centering
	\includegraphics[width=0.32\linewidth]{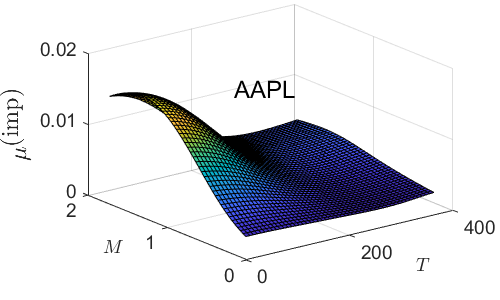}
	\includegraphics[width=0.32\linewidth]{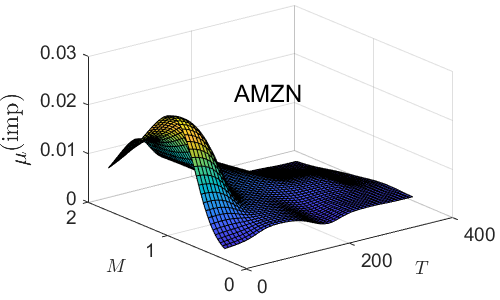}
	\includegraphics[width=0.32\linewidth]{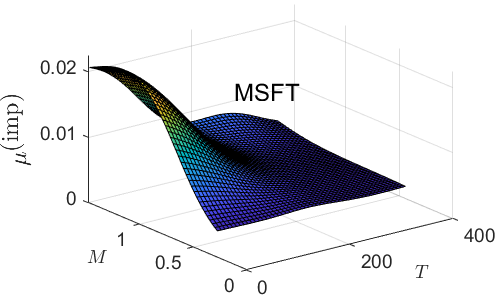}
	\includegraphics[width=0.32\linewidth]{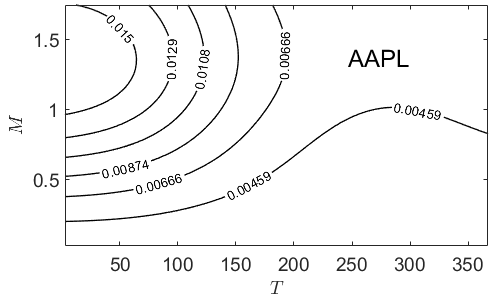}
	\includegraphics[width=0.32\linewidth]{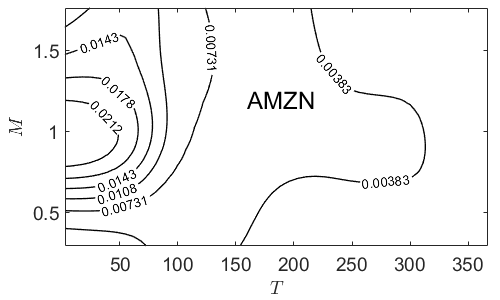}
	\includegraphics[width=0.32\linewidth]{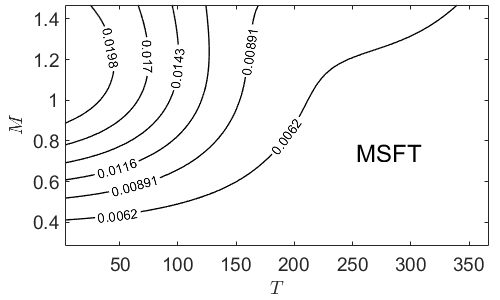}
	\caption{(top row) Computed implied mean surfaces $\mu^{(\text{imp})}(T,M)$
		for call options for the three assets.
		(bottom row) Surface contours projected on the ($T,M$) plane.
	}
	 \label{fig:IM}
\end{figure}

The implied mean is given by
\begin{equation}\label{eq:IM}
\mu^{(\text{imp})}(T_i,K_j) = \argmin_\mu
	 \left (
		\frac{G^{(\text{th})} \left( S_0,T_i,K_j;\sigma_t^{(r)}, \mu , r_{\text{thr}}, r_{f,t} \right) - G^{(\text{emp})}(S_0,T_i,K_j) }
			{G^{(\text{emp})}(S_0,T_i,K_j)}
	\right )^2 .
\end{equation}
The resultant implied mean surfaces, and projected contours,
for call options are shown in Fig.~\ref{fig:IM}.
For all three stocks, the contour levels are higher than the respective historical value of
$\mu$ presented in Table~\ref{tab:params}.
Thus for all projected $T$ and $M$, (on 01/16/2024) the option traders projected returns for these
three stocks greater than the historical return.

\subsection{Implied risk-free rate} \label{sec:IRf}
\begin{figure}[htbp]
	\centering
	\includegraphics[width=0.32\linewidth]{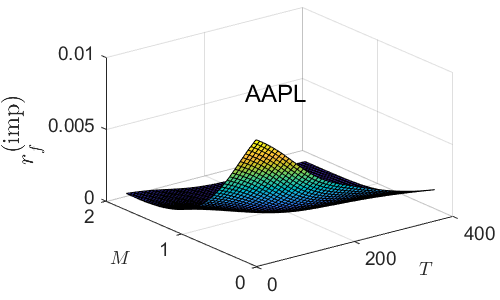}
	\includegraphics[width=0.32\linewidth]{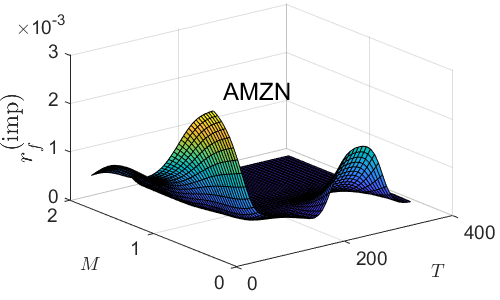}
	\includegraphics[width=0.32\linewidth]{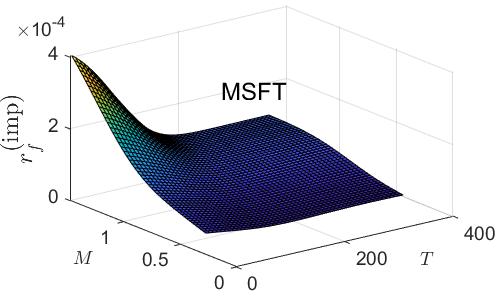}
	\includegraphics[width=0.32\linewidth]{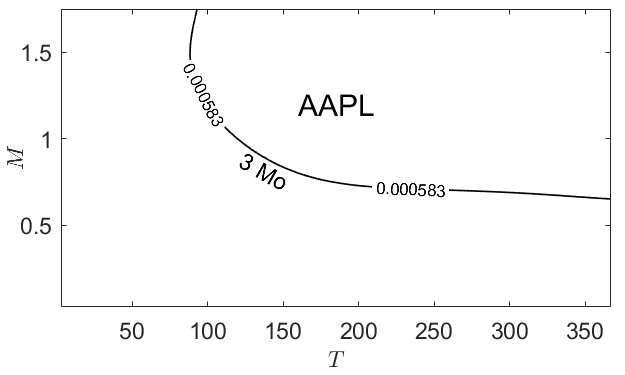}
	\includegraphics[width=0.32\linewidth]{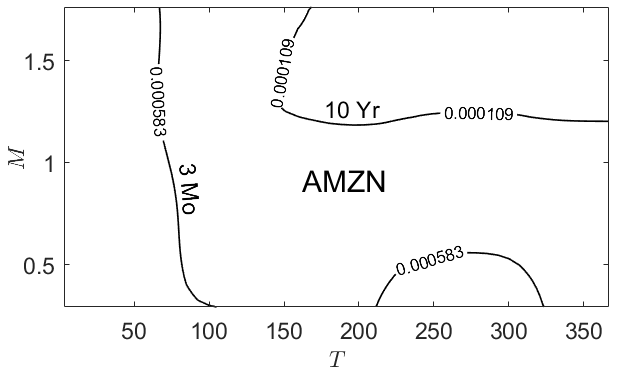}
	\includegraphics[width=0.32\linewidth]{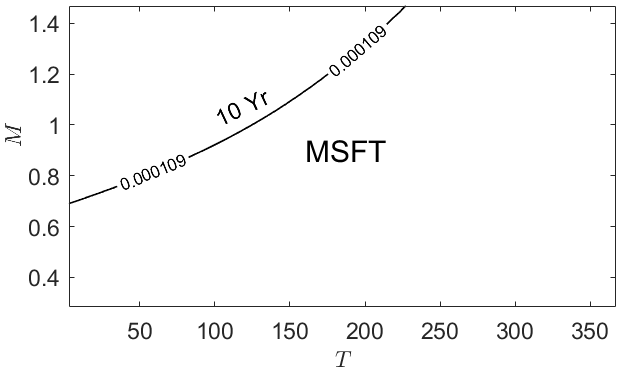}
	\caption{(top row) Computed implied risk-free rate surfaces $r_f^{(\text{imp})}(T,M)$
		for call options for the three assets.
		(bottom row) Surface contours corresponding to the historical three month and
		10 year riskfree rate projected on the ($T,M$) plane.
	}
	 \label{fig:IRF}
\end{figure}

The implied risk-free rate is given by
\begin{equation}\label{eq:IP}
r_f^{(\text{imp})}(T_i,K_j) = \argmin_{r_f}
	 \left (
		\frac{G^{(\text{th})} \left( S_0,T_i,K_j;\sigma_t^{(r)}, \mu_t^{(r)} , r_{\text{thr}}, r_f \right) - G^{(\text{emp})}(S_0,T_i,K_j) }
			{G^{(\text{emp})}(S_0,T_i,K_j)}
	\right )^2 .
\end{equation}
The resultant implied risk-free rate surfaces, and projected contours, for call options are shown in
Fig.~\ref{fig:IRF}.
In this case only two contour levels are drawn, corresponding to the 10 yr. and 3 mo.
risk-free rate values in Table~\ref{tab:params}.
For AAPL, all values of $r_f^{\textrm{(imp)}}$ exceed the 10 yr rate.
However the 3 mo. risk-free rate contour separates the $(T,M)$ plane into two pieces.
Thus, while (on 01/16/2024) option traders viewed future investment in AAPL a superior to investing
in a risk-free 10-year bond,
there is a split on whether to invest in AAPL versus a three month treasury bill.
The region ``below and left'' of the contour favors investment in AAPL,
while the region ``above and right'' favors investment in the Treasury bill.
For AMZN, both the 10 yr. and 3 mo. contours appear indicating $(T,M)$ dependence on lack of confidence
in investing in AMZN compared to the Treasury bill or bond.
For MSFT, only the 10 yr. contour appears.
Thus option traders favored investment in the 3 mo. bill over MSFT,
while there is a regional $(T,M)$ split as to whether to
invest in MSFT compared to the 10 yr. bond.

\subsection{Implied price change probability} \label{sec:IP}
\begin{figure}[htbp]
	\centering
	\includegraphics[width=0.32\linewidth]{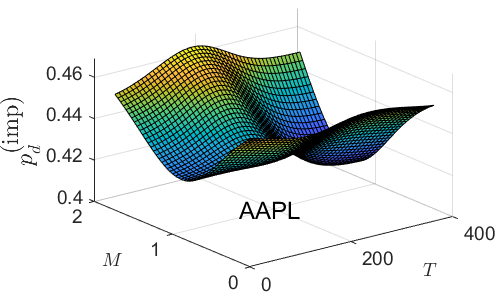}
	\includegraphics[width=0.32\linewidth]{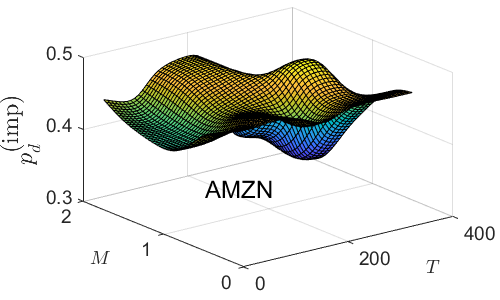}
	\includegraphics[width=0.32\linewidth]{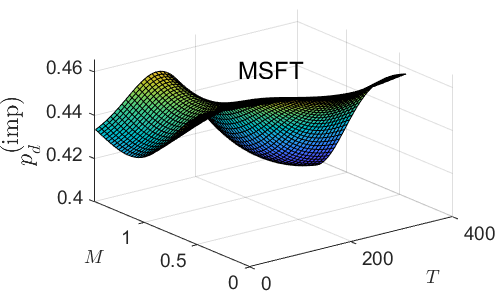}
	\includegraphics[width=0.32\linewidth]{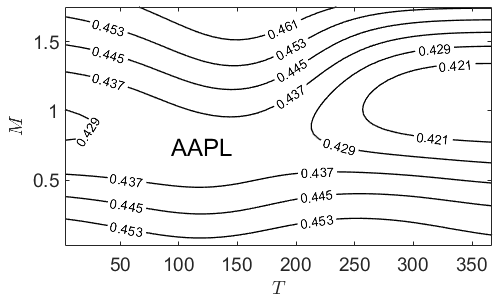}
	\includegraphics[width=0.32\linewidth]{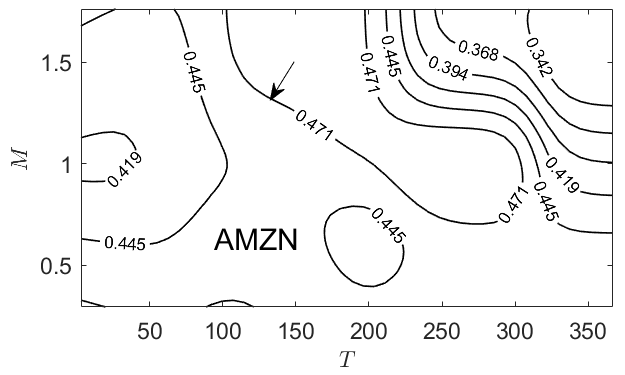}
	\includegraphics[width=0.32\linewidth]{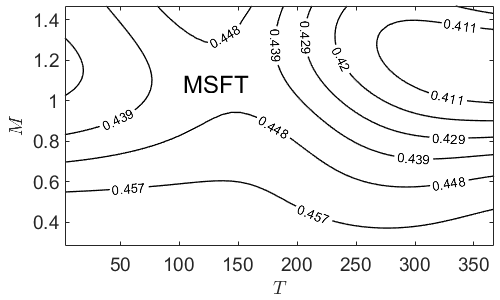}
	\caption{(top row) Computed implied probability $p_d^{\text{(imp)}} (T,M)$
		surfaces for call options for the three assets.
		(bottom row) Surface contours projected on the ($T,M$) plane.
		Arrows indicate contour closest to historical value of $p_d$.
	}
	 \label{fig:IPd}
\end{figure}

\begin{figure}[htbp]
	\centering
	\includegraphics[width=0.32\linewidth]{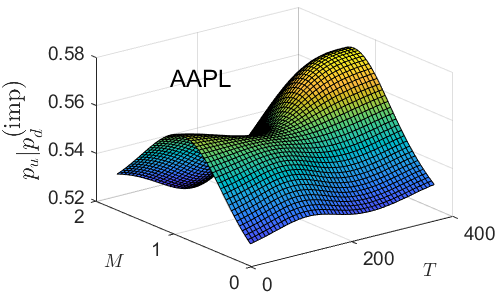}
	\includegraphics[width=0.32\linewidth]{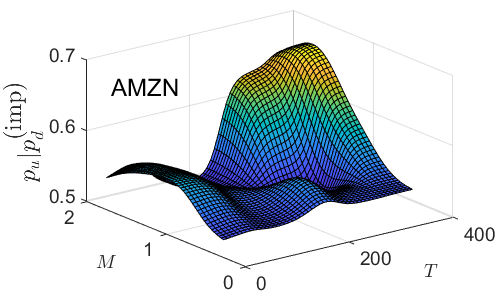}
	\includegraphics[width=0.32\linewidth]{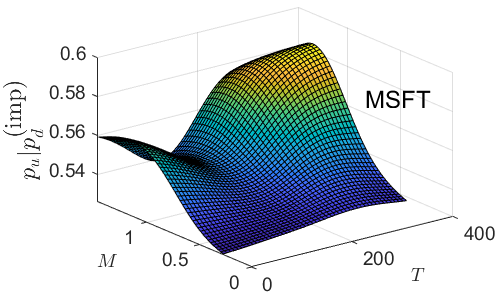}
	\includegraphics[width=0.32\linewidth]{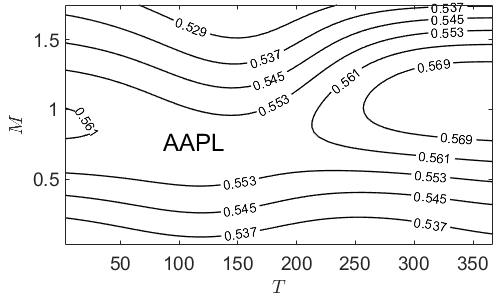}
	\includegraphics[width=0.32\linewidth]{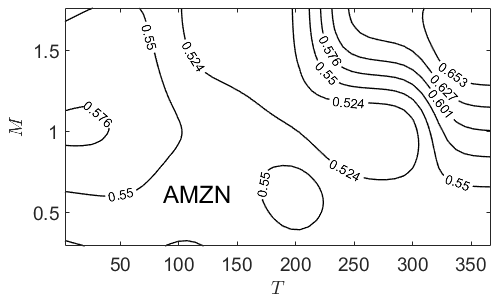}
	\includegraphics[width=0.32\linewidth]{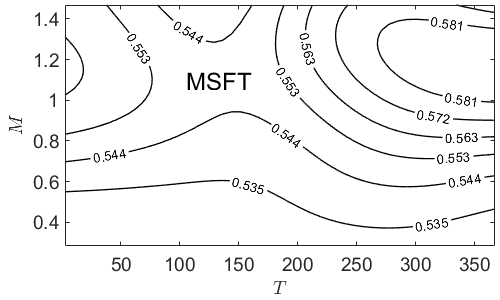}
	\caption{(top row) The surfaces $p_u|p_d^{(\text{imp})}(T,M)$.
		(bottom row) Surface contours projected on the ($T,M$) plane.
	}
	 \label{fig:PuIPd}
\end{figure}

For the implied parameters $\sigma$, $\mu$, and $r_f$,
specification of $r_{\text{thr}}$ completely determines $p_{u,t}$, $p_{m,t}$ and $p_{d,t}$
used on the trinomial tree.
However, we now wish to compute implied probabilities for each pair $T_i,K_j$.
As noted in Section~\ref{sec:r_ext}, we consider the computation of implied values for $p_d$
by holding $p_m$ constant (to the appropriate stock value $p_{m,t}^{(r)}$ given in Table~\ref{tab:params})
and requiring $0 \le p_d \le 1-p_{m,t}^{(r)}$.
Thus the implied probability for $p_d$ is given by
\begin{equation}\label{eq:IPd}
p_d^{(\text{imp})}(T_i,K_j) = \argmin_{0 \le p_d \le 1-p_{m,t}^{(r)}}
	 \left (
		\frac{G^{(\text{th})} \left( S_0,T_i,K_j;\sigma_t^{(r)}, \mu_t^{(r)} , p_{m,t}^{(r)}, p_d, r_{f,t} \right)
			- G^{(\text{emp})}(S_0,T_i,K_j) }
			{G^{(\text{emp})}(S_0,T_i,K_j)}
	\right )^2 .
\end{equation}
Using the value $p_{m,t}^{(r)}$ and the implied values $p_d^{(\text{imp})}(T_i,K_j)$,
we can compute the surface of values
$$
	p_u | p_d^{(\text{imp})} (T_i,K_j) = 1  -p_{m,t}^{(r)} - p_d^{(\text{imp})}(T_i,K_j).
$$
Fig.~\ref{fig:IPd} plots the $p_d^{\text{(imp)}} (T,M)$ surfaces, and projected contours,
for the three stocks.
Compared to the range of $p_d^{\text{(imp)}}$ values evidenced for AMZN,
those for AAPL and MSFT are, essentially flat.
In addition, all coutour levels for AAPL and MSFT fall below respective the historical value of $p_d$.
For these two stocks, option traders projected a (slightly) decreased probability for a price downturn
than that given by the historical  value.
For AMZN, the historical value of $p_d = 0.477$ is very close to the 0.471 contour level.
While option traders saw a much larger range of projected price downturn probabilities,
there is a (smaller) $(T,M)$ region where they projected higher probability for price downturn,
as well as the larger complement region projecting lower probability for the same.
Using the value $p_{m,t}^{(r)}$ and the implied values $p_d^{(\text{imp})}(T_i,K_j)$,
we can compute the values
$$
	p_u | p_d^{(\text{imp})} (T_i,K_j) = 1  -p_{m,t}^{(r)} - p_d^{(\text{imp})}(T_i,K_j).
$$
Fig.~\ref{fig:PuIPd} plots the $p_u | p_d^{\text{(imp)}} (T,M)$ surfaces, and projected contours,
for the three stocks.
As $p_d^{\text{(imp)}} (T,M) + p_u | p_d^{\text{(imp)}} (T,M) = 1 - p_m$ (a constant),
there is no additional information in these plots.
For AAPL and MSFT, option traders projected a (slightly) increased probability for a price increase
than that given by the historical  value.
For MSFT option traders saw a much larger range of projected price increase probabilities,
with (most of the) region corresponding to probability larger than the historical value.

\begin{figure}[htbp]
	\centering
	\includegraphics[width=0.32\linewidth]{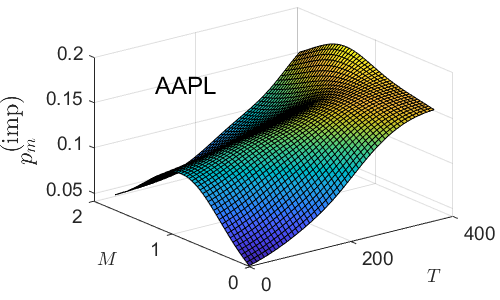}
	\includegraphics[width=0.32\linewidth]{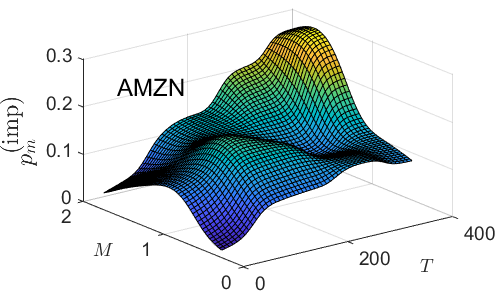}
	\includegraphics[width=0.32\linewidth]{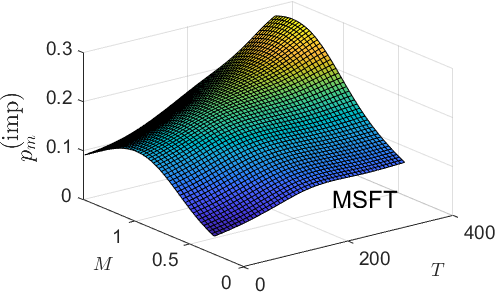}
	\includegraphics[width=0.32\linewidth]{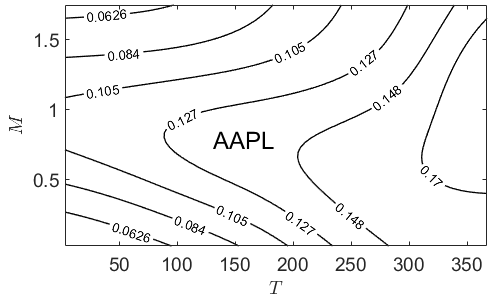}
	\includegraphics[width=0.32\linewidth]{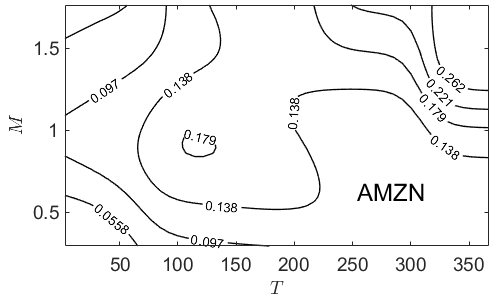}
	\includegraphics[width=0.32\linewidth]{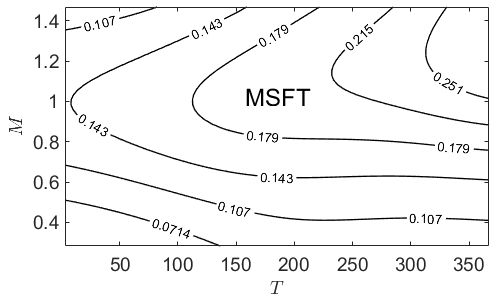}
	\caption{(top row) Computed implied probability $p_m^{\text{(imp)}} (T,M)$
		surfaces for call options for the three assets.
		(bottom row) Surface contours projected on the ($T,M$) plane.
	}
	 \label{fig:IPm}
\end{figure}

\begin{figure}[htbp]
	\centering
	\includegraphics[width=0.32\linewidth]{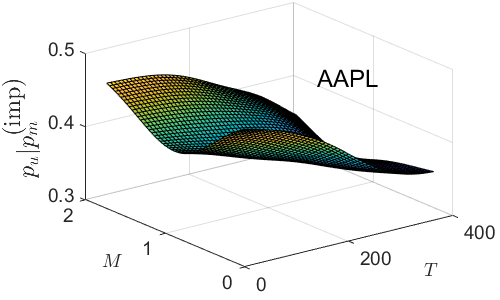}
	\includegraphics[width=0.32\linewidth]{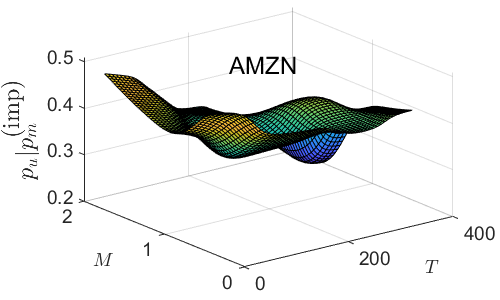}
	\includegraphics[width=0.32\linewidth]{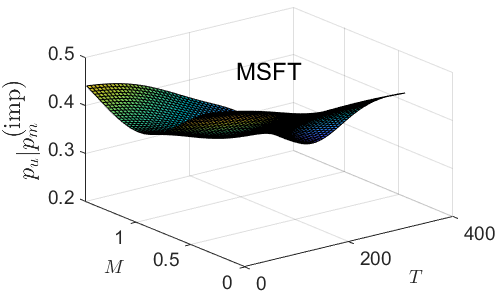}
	\includegraphics[width=0.32\linewidth]{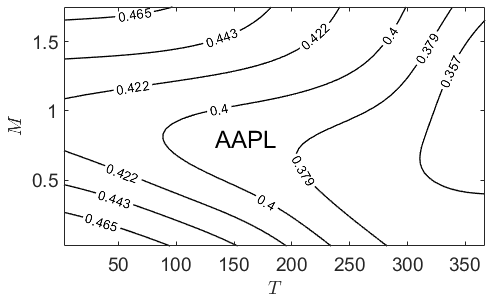}
	\includegraphics[width=0.32\linewidth]{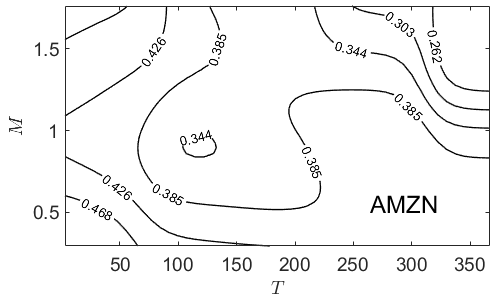}
	\includegraphics[width=0.32\linewidth]{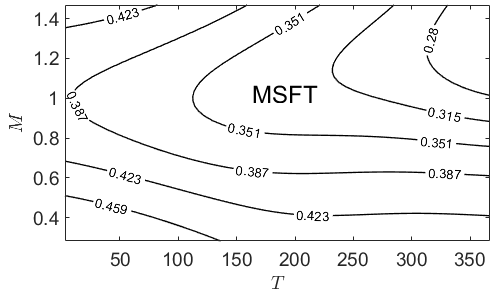}
	\caption{(top row) The surfaces $p_u|p_m^{(\text{imp})}(T,M)$.
		(bottom row) Surface contours projected on the ($T,M$) plane.
	}
	 \label{fig:PuIPm}
\end{figure}

Similarly, implied values for $p_m$ are computed via
\begin{equation}\label{eq:IPd}
p_m^{(\text{imp})}(T_i,K_j) = \argmin_{0 \le p_m \le 1-p_{d,t}^{(r)}}
	 \left (
		\frac{G^{(\text{th})} \left( S_0,T_i,K_j;\sigma_t^{(r)}, \mu_t^{(r)} , p_{d,t}^{(r)}, p_m, r_{f,t} \right)
			- G^{(\text{emp})}(S_0,T_i,K_j) }
			{G^{(\text{emp})}(S_0,T_i,K_j)}
	\right )^2 ,
\end{equation}
from which we can compute the values
$$
	p_u | p_m^{(\text{imp})} (T_i,K_j) = 1  - p_{d,t}^{(r)} - p_m^{(\text{imp})}(T_i,K_j).
$$
Fig.~\ref{fig:IPm} plots the $p_m^{\text{(imp)}} (T,M)$ surfaces and contours.
In contrast to the $p_d^{\text{(imp)}}$ surfaces, the range of values for $p_m^{\text{(imp)}}$
is large for all three stocks.
All contour levels are greater (by at minimum 5 to 10 times) than the respective historical values
of $p_m$ in Table~\ref{tab:params}.
And the larger the value of implied $p_m$, the less confidence an option trader places on whether
the stock price will go up\footnote{
	It is important to keep in mind that the implied values of $p_m$ are computed assuming the
	probability for a price downturn $p_d$ is fixed at the historical  value.
}.
Thus large maturity time, far out-of-the-money option prices correspond to the largest values for
$p_m^{\text{(imp)}}$.
And $p_m^{\text{(imp)}}$ values should generally increase as T increases.
Again, for completeness,
Fig.~\ref{fig:PuIPm}  plots the $p_u | p_m^{(\text{imp})}(T,M)$ surfaces and contours.
As a fixed value for $p_d$ is used in the $p_m^{(\text{imp})}$ computations,
no additional information is available in these plots.
The large range of values for $p_m^{(\text{imp})}$ lead to very conservative projections for the values of the probability $p_u$,
all of which fall below the historical values presented in Table~\ref{tab:params}.

\subsection{Implied extreme price change probability} \label{sec:IPext}
\begin{figure}[htbp]
	\centering
	\includegraphics[width=0.32\linewidth]{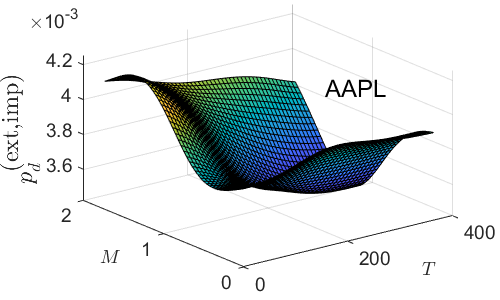}
	\includegraphics[width=0.32\linewidth]{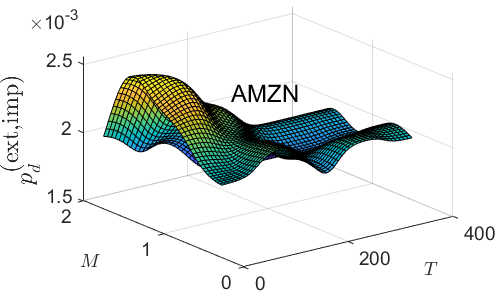}
	\includegraphics[width=0.32\linewidth]{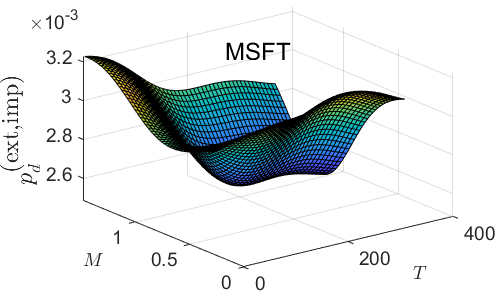}
	\includegraphics[width=0.32\linewidth]{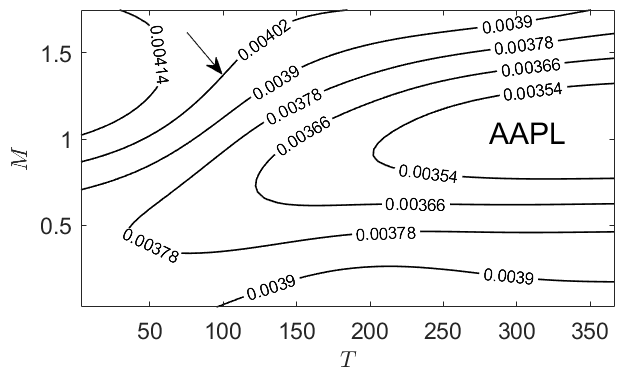}
	\includegraphics[width=0.32\linewidth]{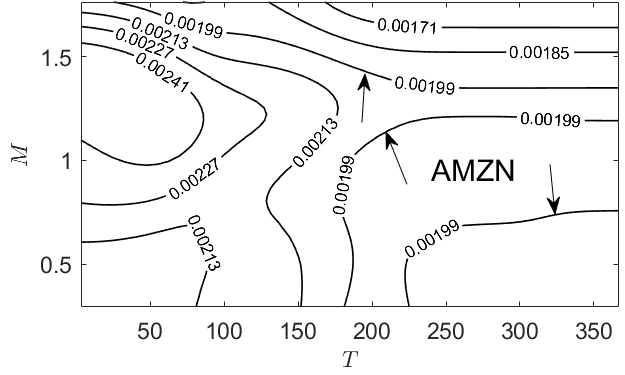}
	\includegraphics[width=0.32\linewidth]{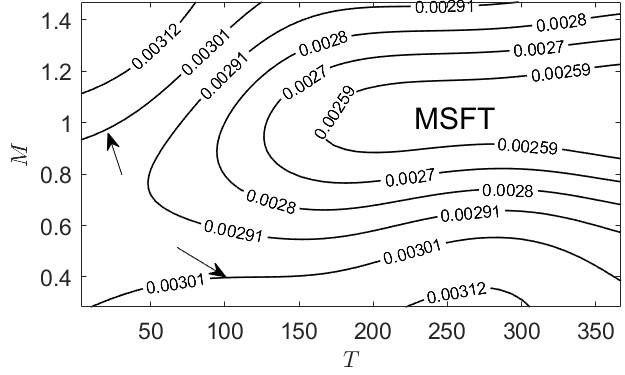}
	\caption{(top row) Computed implied probability $p_d^{\text{(ext ,imp)}} (T,M)$
		surfaces for call options for the three assets.
		(bottom row) Surface contours projected on the ($T,M$) plane.
		Arrows indicate contour closest to historical value of $p_d^{\text{(ext )}}$.
	}
	 \label{fig:IPdext}
\end{figure}

\begin{table}[htb]
	\caption{Price change probabilities computed from  $\text{CVaR}_{0.01}$ and  $\overline{\text{CVaR}}_{0.01}$}
	\label{tab:IPext}
	\centering
	\begin{tabular}{l cc ccc}
	\toprule
	Stock   & $\text{CVaR}_{0.01}$  &$ \overline{\text{CVaR}}_{0.01}$  &
				$p_d^{\text{(ext)}}$ & $p_m^{\text{(ext)}}$ & $p_u^{\text{(ext)}}$ \\
	\midrule
	AAPL    &  $-0.0754$ &  $0.0875$ & $0.00398$ & 0.991 &  $0.00498$\\
	AMZN  &  $-0.0820$ &  $0.0874$ & $0.00199$ & 0.995 &  $0.00298$\\
	MSFT   &  $-0.0733$ &  $0.0817$ & $0.00298$ & 0.993 &  $0.00398$\\
	\bottomrule
	\end{tabular}
\end{table}
Table~\ref{tab:IPext} lists the one percent conditional value-at-risk and conditional value-at-return,
as well as the resultant extreme price change values, computed from the historical return data. Fig.~\ref{fig:IPdext} plots the implied surface $p_d^{\text{(ext, imp)}} (T,M)$,
as well as projected surface contours,
computed holding $p_m^{\text{(ext)}}$ at the historical value. For AAPL the historical value of $p_d^{\text{(ext)}}$ is approximated by the 0.00402 contour;
for AMZN the historical $p_d^{\text{(ext)}}$ corresponds to the 0.00199 contour;
and for MSFT it is approximated by the 0.00301 contour.
\begin{figure}[htbp]
	\centering
	\includegraphics[width=0.32\linewidth]{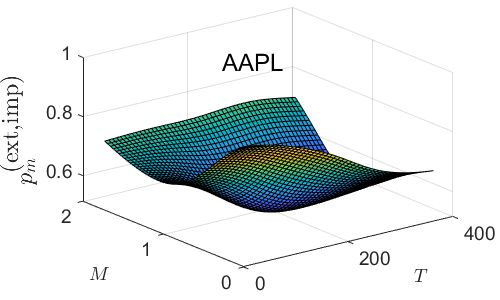}
	\includegraphics[width=0.32\linewidth]{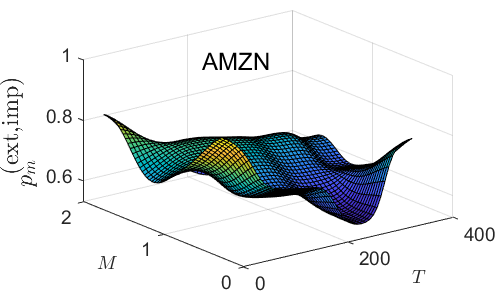}
	\includegraphics[width=0.32\linewidth]{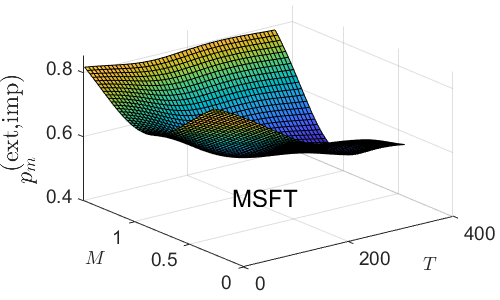}
	\includegraphics[width=0.32\linewidth]{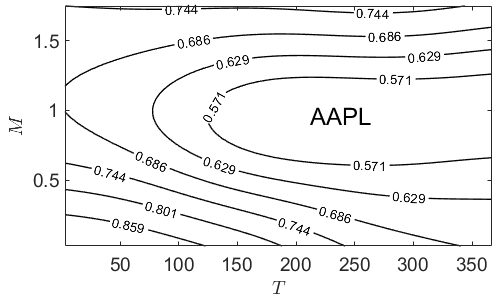}
	\includegraphics[width=0.32\linewidth]{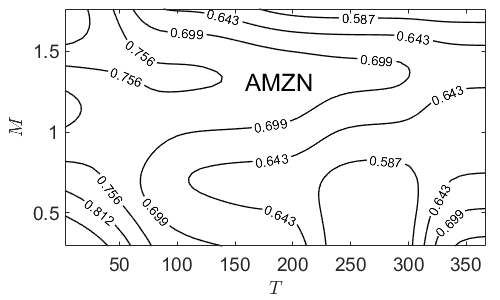}
	\includegraphics[width=0.32\linewidth]{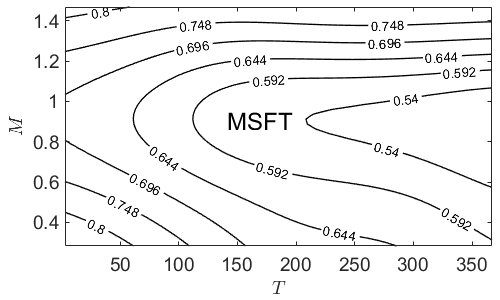}
	\caption{(top row) Computed implied probability $p_m^{\text{(ext, imp)}} (T,M)$
		surfaces for call options for the three assets.
		(bottom row) Surface contours projected on the ($T,M$) plane.
	}
	 \label{fig:IPmext}
\end{figure}

Thus, with small variation, for all three stocks option traders view the probability of extreme
downward movement in the price roughly similarly to spot traders.
Fig.~\ref{fig:IPmext} plots the surface $p_m^{\text{(ext, imp)}} (T,M)$,
as well as projected surface contours,
computed holding $p_d^{\text{(ext)}}$ at the historical value.
For all three stocks, the implied values $p_m^{\text{(ext, imp)}}$ fall significantly below
the historical value.
Thus option traders view the probability for  non-extreme movements of the price to be lower than
that of spot traders.


\newpage
\section{Conclusion} \label{sec:Conclusion}
This work makes the following significant contributions to the literature on trinomial models.
\begin{itemize}
\item
We have developed a market complete trinomial pricing model in which completeness is ensured by a market consisting of:
a stock and its perpetual  derivative as risky assets; a riskless asset (bond); and a European option.
The use of the perpetual derivative ensures that the number of Brownian motions driving price stochastics does not increase,
thus ensuring the completeness of the market.
\item
Our model is developed in the natural world and, through the construction of a replicating portfolio,
we derive the risk-neutral price dynamics of all four assets.
This methodology thus captures the relationship between the risk-neutral and natural-world parameters.
\item
We derive a new approach for calibrating the probabilities $p_d, p_m$ and $p_d$ for price movements in the natural-world model  to empirical data.
The approach is based upon hypothesis testing on sub-sample mean values.
\item
As a result of capturing the explicit relationship between the risk-neutral and natural-world parameters,
using call option data from three of the ``Magnificant Seven'' technology stocks
we compute implied surfaces for all parameters in the model.
Examination of the contour levels of an implied parameter surface may split the surface into two regimes
 -- ``above and below'' the historical value for that parameter --
allowing for a comparison of the views of option and spot traders relative to the future performance of that parameter.
\end{itemize}
\renewcommand{\thetable}{A.\arabic{table}}
\setcounter{table}{0}


\begin{appendices}
\section{Determination of $\rthr$ Values} \label {secA1}
With reference to the discussion in Section~\ref{sec:r_est},
Table~\ref{tab:rthr} shows how $\rthr^-$ and $\rthr^+$ vary as a function of the significance level
$\alpha \in \{0.05, 0.01, 0.005, 0.001\}$ based on the historical window of returns observed for the
three indicated stocks.
Here $p_{J^-}$ indicates the smallest value of $p_j = j \delta p < 0$ for which the null hypothesis
$H_0 : \mu_{p_j}$ is not rejected at the significance level $\alpha$, 
while $p_{J^+}$ indicates the largest value of $p_j = j \delta p > 0$ for which the null hypothesis
$H_0 : \mu_{p_j}$ is not rejected.
The computations were done with $\delta p = \pm 1$ basis point.

\begin{table}[htb]
	\caption{$\rthr$ values}
	\label{tab:rthr}
	\centering
	\begin{tabular}{l c lclc}
	\toprule
	Stock & $\alpha$	& $p_{J^-}$	& $\rthr^-$			& $p_{J^+}$	& $\rthr^+$ \\
	\midrule
	AAPL & 0.050		& $-$3		& $-7.48 \cdot 10^{-5}$	& NS		&NS				      \\
	\omit & 0.010		& $-$3		& $-7.48 \cdot 10^{-5}$	& 1 			&$3.88 \cdot 10^{-5}$  \\
	\omit & 0.005		& $-$3		& $-7.48 \cdot 10^{-5}$	& 3			& $7.96 \cdot 10^{-5}$ \\
	\omit & 0.001		& $-$5		& $-2.06 \cdot 10^{-4}$	& 4			& $1.46 \cdot 10^{-4}$ 
	\vspace{0.1in} \\
	AMZN & 0.050		& $-$1		& $-2.55 \cdot 10^{-5}$	& 1 			& 0  \\
	\omit & 0.010		& $-$2		& $-7.74 \cdot 10^{-5}$	& 1 			& 0  \\
	\omit & 0.005		& $-$2		& $-7.74 \cdot 10^{-5}$	& 2			& $9.83 \cdot 10^{-5}$ \\
	\omit & 0.001		& $-$3		& $-1.26 \cdot 10^{-4}$	& 2			& $9.83 \cdot 10^{-5}$
	\vspace{0.1in} \\
	MSFT & 0.050		& $-$1		& $-1.50 \cdot 10^{-5}$	& NS		& NS  \\
	\omit & 0.010		& $-$2		& $-7.65 \cdot 10^{-5}$	& 1 			&$2.87 \cdot 10^{-5}$ \\
	\omit & 0.005		& $-$2		& $-7.65 \cdot 10^{-5}$	& 2			& $6.89 \cdot 10^{-5}$ \\
	\omit & 0.001		& $-$3		& $-1.06 \cdot 10^{-4}$	& 4			& $1.16 \cdot 10^{-4}$ \\
	\bottomrule
	\multicolumn{6}{l}{NS indicates the null hypothesis was rejected for all values}\\
	\multicolumn{6}{l}{$p_j = j \delta p$, $j = 1, 2, ...\  $.}
	\end{tabular}
\end{table}
\vspace{5cm}

\end{appendices}

\bibliography{OP_tri_tree}


\begin{thebibliography}{31}
\ifx \bisbn   \undefined \def \bisbn  #1{ISBN #1}\fi
\ifx \binits  \undefined \def \binits#1{#1}\fi
\ifx \bauthor  \undefined \def \bauthor#1{#1}\fi
\ifx \batitle  \undefined \def \batitle#1{#1}\fi
\ifx \bjtitle  \undefined \def \bjtitle#1{#1}\fi
\ifx \bvolume  \undefined \def \bvolume#1{\textbf{#1}}\fi
\ifx \byear  \undefined \def \byear#1{#1}\fi
\ifx \bissue  \undefined \def \bissue#1{#1}\fi
\ifx \bfpage  \undefined \def \bfpage#1{#1}\fi
\ifx \blpage  \undefined \def \blpage #1{#1}\fi
\ifx \burl  \undefined \def \burl#1{\textsf{#1}}\fi
\ifx \doiurl  \undefined \def \doiurl#1{\url{https://doi.org/#1}}\fi
\ifx \betal  \undefined \def \betal{\textit{et al.}}\fi
\ifx \binstitute  \undefined \def \binstitute#1{#1}\fi
\ifx \binstitutionaled  \undefined \def \binstitutionaled#1{#1}\fi
\ifx \bctitle  \undefined \def \bctitle#1{#1}\fi
\ifx \beditor  \undefined \def \beditor#1{#1}\fi
\ifx \bpublisher  \undefined \def \bpublisher#1{#1}\fi
\ifx \bbtitle  \undefined \def \bbtitle#1{#1}\fi
\ifx \bedition  \undefined \def \bedition#1{#1}\fi
\ifx \bseriesno  \undefined \def \bseriesno#1{#1}\fi
\ifx \blocation  \undefined \def \blocation#1{#1}\fi
\ifx \bsertitle  \undefined \def \bsertitle#1{#1}\fi
\ifx \bsnm \undefined \def \bsnm#1{#1}\fi
\ifx \bsuffix \undefined \def \bsuffix#1{#1}\fi
\ifx \bparticle \undefined \def \bparticle#1{#1}\fi
\ifx \barticle \undefined \def \barticle#1{#1}\fi
\bibcommenthead
\ifx \bconfdate \undefined \def \bconfdate #1{#1}\fi
\ifx \botherref \undefined \def \botherref #1{#1}\fi
\ifx \url \undefined \def \url#1{\textsf{#1}}\fi
\ifx \bchapter \undefined \def \bchapter#1{#1}\fi
\ifx \bbook \undefined \def \bbook#1{#1}\fi
\ifx \bcomment \undefined \def \bcomment#1{#1}\fi
\ifx \oauthor \undefined \def \oauthor#1{#1}\fi
\ifx \citeauthoryear \undefined \def \citeauthoryear#1{#1}\fi
\ifx \endbibitem  \undefined \def \endbibitem {}\fi
\ifx \bconflocation  \undefined \def \bconflocation#1{#1}\fi
\ifx \arxivurl  \undefined \def \arxivurl#1{\textsf{#1}}\fi
\csname PreBibitemsHook\endcsname

\bibitem[\protect\citeauthoryear{Ahn and Song}{2007}]{ahn_2007}
\begin{barticle}
\bauthor{\bsnm{Ahn}, \binits{J.}},
\bauthor{\bsnm{Song}, \binits{M.}}:
\batitle{Convergence of the trinomial tree method for pricing {E}uropean/{A}merican options}.
\bjtitle{Applied Mathematics and Computation}
\bvolume{189}(\bissue{1}),
\bfpage{575}--\blpage{582}
(\byear{2007})
\end{barticle}
\endbibitem

\bibitem[\protect\citeauthoryear{Bates}{1996}]{bates_1996}
\begin{barticle}
\bauthor{\bsnm{Bates}, \binits{D.S.}}:
\batitle{Jumps and stochastic volatility: {E}xchange rate processes implicit in {D}eutsche {M}ark options}.
\bjtitle{Review of Financial Studies}
\bvolume{9},
\bfpage{69}--\blpage{108}
(\byear{1996})
\end{barticle}
\endbibitem

\bibitem[\protect\citeauthoryear{Boyle et~al.}{1989}]{boyle_1989}
\begin{barticle}
\bauthor{\bsnm{Boyle}, \binits{P.P.}},
\bauthor{\bsnm{Evnine}, \binits{J.}},
\bauthor{\bsnm{Gibbs}, \binits{S.}}:
\batitle{Numerical evaluation of multivariate contingent claims}.
\bjtitle{Review of Financial Studies}
\bvolume{2},
\bfpage{241}--\blpage{251}
(\byear{1989})
\end{barticle}
\endbibitem

\bibitem[\protect\citeauthoryear{Boyle and Lau}{1994}]{boyle_1994}
\begin{barticle}
\bauthor{\bsnm{Boyle}, \binits{P.P.}},
\bauthor{\bsnm{Lau}, \binits{S.H.}}:
\batitle{Bumping up against the barrier with the binomial method}.
\bjtitle{Journal of Derivatives}
\bvolume{1},
\bfpage{6}--\blpage{14}
(\byear{1994})
\end{barticle}
\endbibitem

\bibitem[\protect\citeauthoryear{Boyle}{1986}]{boyle_1986}
\begin{barticle}
\bauthor{\bsnm{Boyle}, \binits{P.P.}}:
\batitle{Option valuation using a tree-jump process}.
\bjtitle{International Options Journal}
\bvolume{3},
\bfpage{7}--\blpage{12}
(\byear{1986})
\end{barticle}
\endbibitem

\bibitem[\protect\citeauthoryear{Boyle}{1988}]{boyle_1988}
\begin{barticle}
\bauthor{\bsnm{Boyle}, \binits{P.P.}}:
\batitle{A lattice framework for option pricing with two state variables}.
\bjtitle{Journal of Financial and Quantitative Analysis}
\bvolume{23}(\bissue{1}),
\bfpage{1}--\blpage{12}
(\byear{1988})
\end{barticle}
\endbibitem

\bibitem[\protect\citeauthoryear{Black and Scholes}{1973}]{black_1973}
\begin{barticle}
\bauthor{\bsnm{Black}, \binits{F.}},
\bauthor{\bsnm{Scholes}, \binits{M.}}:
\batitle{The pricing of options and corporate liabilities}.
\bjtitle{Journal of Political Economy}
\bvolume{81}(\bissue{3}),
\bfpage{637}--\blpage{654}
(\byear{1973})
\end{barticle}
\endbibitem

\bibitem[\protect\citeauthoryear{Chan et~al.}{2009}]{Chan_2009}
\begin{barticle}
\bauthor{\bsnm{Chan}, \binits{J.H.}},
\bauthor{\bsnm{Joshi}, \binits{M.}},
\bauthor{\bsnm{Tang}, \binits{R.}},
\bauthor{\bsnm{Yang}, \binits{C.}}:
\batitle{Trinomial or binomial: {A}ccelerating {A}merican put option price on trees}.
\bjtitle{Journal of Futures Markets}
\bvolume{29}(\bissue{9}),
\bfpage{797}--\blpage{893}
(\byear{2009})
\end{barticle}
\endbibitem

\bibitem[\protect\citeauthoryear{Cox et~al.}{1979}]{cox_1979}
\begin{barticle}
\bauthor{\bsnm{Cox}, \binits{J.C.}},
\bauthor{\bsnm{Ross}, \binits{S.A.}},
\bauthor{\bsnm{Rubinstein}, \binits{M.}}:
\batitle{Option pricing: {A} simplified approach}.
\bjtitle{Journal of Financial Economics}
\bvolume{7}(\bissue{3}),
\bfpage{229}--\blpage{263}
(\byear{1979})
\end{barticle}
\endbibitem

\bibitem[\protect\citeauthoryear{Deutsch}{2009}]{deutsch_2009}
\begin{bbook}
\bauthor{\bsnm{Deutsch}, \binits{H.-P.}}:
\bbtitle{Derivatives and Internal Models},
\bedition{4}th edn.
\bpublisher{Palgrave Macmillan},
\blocation{New York}
(\byear{2009})
\end{bbook}
\endbibitem

\bibitem[\protect\citeauthoryear{Davydov and Rotar}{2008}]{Davydov_2008}
\begin{barticle}
\bauthor{\bsnm{Davydov}, \binits{Y.}},
\bauthor{\bsnm{Rotar}, \binits{V.}}:
\batitle{On a non-classical invariance principle}.
\bjtitle{Statistics \& Probability Letters}
\bvolume{78},
\bfpage{2031}--\blpage{2038}
(\byear{2008})
\end{barticle}
\endbibitem

\bibitem[\protect\citeauthoryear{Florescu and Viens}{2008}]{florescu_2008}
\begin{barticle}
\bauthor{\bsnm{Florescu}, \binits{I.}},
\bauthor{\bsnm{Viens}, \binits{F.}}:
\batitle{Stochastic volatility: {O}ption pricing using a multinomial recombining tree}.
\bjtitle{Applied Mathematical Finance}
\bvolume{15},
\bfpage{151}--\blpage{181}
(\byear{2008})
\end{barticle}
\endbibitem

\bibitem[\protect\citeauthoryear{Hilliard and Schwartz}{1996}]{hilliard_1996}
\begin{barticle}
\bauthor{\bsnm{Hilliard}, \binits{J.E.}},
\bauthor{\bsnm{Schwartz}, \binits{A.}}:
\batitle{Binomial option pricing under stochastic volatility and correlated state variables}.
\bjtitle{Journal of Derivatives}
\bvolume{4}(\bissue{1}),
\bfpage{23}--\blpage{39}
(\byear{1996})
\end{barticle}
\endbibitem

\bibitem[\protect\citeauthoryear{Hu et~al.}{2020}]{hu_2020}
\begin{barticle}
\bauthor{\bsnm{Hu}, \binits{Y.}},
\bauthor{\bsnm{Shirvani}, \binits{A.}},
\bauthor{\bsnm{Lindquist}, \binits{W.B.}},
\bauthor{\bsnm{Fabozzi}, \binits{F.J.}},
\bauthor{\bsnm{Rachev}, \binits{S.T.}}:
\batitle{Option pricing incorporating factor dynamics in complete markets}.
\bjtitle{Journal of Risk and Financial Management}
\bvolume{13}(\bissue{12}),
\bfpage{321}
(\byear{2020})
\end{barticle}
\endbibitem

\bibitem[\protect\citeauthoryear{Josheski and Apostolov}{2020}]{josheski_2020}
\begin{barticle}
\bauthor{\bsnm{Josheski}, \binits{D.}},
\bauthor{\bsnm{Apostolov}, \binits{M.}}:
\batitle{A review of the binomial and trinomial models for option pricing and their convergence to the {B}lack-{S}choles model determined option prices}.
\bjtitle{Econometrics}
\bvolume{24}(\bissue{2}),
\bfpage{53}--\blpage{85}
(\byear{2020})
\end{barticle}
\endbibitem

\bibitem[\protect\citeauthoryear{Jarrow and Rudd}{1983}]{jarrow_1983}
\begin{bbook}
\bauthor{\bsnm{Jarrow}, \binits{R.}},
\bauthor{\bsnm{Rudd}, \binits{A.}}:
\bbtitle{Option Pricing}.
\bpublisher{Dow Jones-Irwin},
\blocation{Homewood, IL}
(\byear{1983})
\end{bbook}
\endbibitem

\bibitem[\protect\citeauthoryear{Kamrad and Ritchen}{1991}]{kamrad_1991}
\begin{barticle}
\bauthor{\bsnm{Kamrad}, \binits{B.}},
\bauthor{\bsnm{Ritchen}, \binits{P.}}:
\batitle{Multinomial approximating models for options with k state variables}.
\bjtitle{Management Science}
\bvolume{37},
\bfpage{1640}--\blpage{1652}
(\byear{1991})
\end{barticle}
\endbibitem

\bibitem[\protect\citeauthoryear{Kim et~al.}{2019}]{kim_2019}
\begin{barticle}
\bauthor{\bsnm{Kim}, \binits{Y.S.}},
\bauthor{\bsnm{Stoyanov}, \binits{S.V.}},
\bauthor{\bsnm{Rachev}, \binits{S.T.}},
\bauthor{\bsnm{Fabozzi}, \binits{F.J.}}:
\batitle{Enhancing binomial and trinomial equity option pricing models}.
\bjtitle{Finance Research Letters}
\bvolume{28},
\bfpage{185}--\blpage{190}
(\byear{2019})
\end{barticle}
\endbibitem

\bibitem[\protect\citeauthoryear{Langat et~al.}{2019}]{langat_2019}
\begin{barticle}
\bauthor{\bsnm{Langat}, \binits{K.K.}},
\bauthor{\bsnm{Mwaniki}, \binits{J.I.}},
\bauthor{\bsnm{Kiprop}, \binits{G.K.}}:
\batitle{Pricing options using trinomial lattice method}.
\bjtitle{Journal of Finance and Economics}
\bvolume{7},
\bfpage{81}--\blpage{87}
(\byear{2019})
\end{barticle}
\endbibitem

\bibitem[\protect\citeauthoryear{Leisen and Reimer}{1996}]{leisen_1996}
\begin{barticle}
\bauthor{\bsnm{Leisen}, \binits{D.P.J.}},
\bauthor{\bsnm{Reimer}, \binits{M.}}:
\batitle{Binomial models for option valuation – examining and improving convergence}.
\bjtitle{Applied Mathematical Finance}
\bvolume{3}(\bissue{4}),
\bfpage{319}--\blpage{346}
(\byear{1996})
\end{barticle}
\endbibitem

\bibitem[\protect\citeauthoryear{Lindquist and Rachev}{2024}]{Lindquist_2024}
\begin{botherref}
\oauthor{\bsnm{Lindquist}, \binits{W.B.}},
\oauthor{\bsnm{Rachev}, \binits{S.T.}}:
Alternatives to classical option pricing.
Annals of Operations Research,
1--21
(2024)
\end{botherref}
\endbibitem

\bibitem[\protect\citeauthoryear{Lilyana et~al.}{2021}]{lilyana_2021}
\begin{bchapter}
\bauthor{\bsnm{Lilyana}, \binits{D.}},
\bauthor{\bsnm{Subartini}, \binits{B.}},
\bauthor{\bsnm{Riaman}, \binits{R.}},
\bauthor{\bsnm{Supriatna}, \binits{A.K.}}:
\bctitle{Calculation of call option using trinomial tree method and {B}lack-{S}choles method: {C}ase study of {M}icrosoft {C}orporation}.
In: \bbtitle{Journal of Physics: Conference Series},
vol. \bseriesno{1722},
p. \bfpage{012064}
(\byear{2021}).
\bcomment{IOP Publishing}
\end{bchapter}
\endbibitem

\bibitem[\protect\citeauthoryear{Merton}{1973}]{merton_1973}
\begin{barticle}
\bauthor{\bsnm{Merton}, \binits{R.C.}}:
\batitle{Theory of rational option pricing}.
\bjtitle{The Bell Journal of Economics and Management Science}
\bvolume{4}(\bissue{1}),
\bfpage{141}--\blpage{183}
(\byear{1973})
\end{barticle}
\endbibitem

\bibitem[\protect\citeauthoryear{Madan et~al.}{1989}]{madan_1989}
\begin{barticle}
\bauthor{\bsnm{Madan}, \binits{D.B.}},
\bauthor{\bsnm{Milne}, \binits{F.}},
\bauthor{\bsnm{Shefrin}, \binits{H.}}:
\batitle{The multinomial option pricing model and its {B}rownian and {P}oisson limits}.
\bjtitle{Review of Financial Studies}
\bvolume{2},
\bfpage{251}--\blpage{265}
(\byear{1989})
\end{barticle}
\endbibitem

\bibitem[\protect\citeauthoryear{Ma and Zhu}{2015}]{ma_2015}
\begin{barticle}
\bauthor{\bsnm{Ma}, \binits{J.}},
\bauthor{\bsnm{Zhu}, \binits{T.}}:
\batitle{Convergence rates of trinomial tree methods for option pricing under regime-switching models}.
\bjtitle{Applied Mathematics Letters}
\bvolume{39},
\bfpage{13}--\blpage{18}
(\byear{2015})
\end{barticle}
\endbibitem

\bibitem[\protect\citeauthoryear{Rubinstein}{1998}]{rubinstein_1998}
\begin{barticle}
\bauthor{\bsnm{Rubinstein}, \binits{M.}}:
\batitle{Edgeworth binomial trees}.
\bjtitle{The Journal of Derivatives}
\bvolume{5}(\bissue{3}),
\bfpage{20}--\blpage{27}
(\byear{1998})
\end{barticle}
\endbibitem

\bibitem[\protect\citeauthoryear{Sharpe}{1978}]{sharpe_1978}
\begin{bbook}
\bauthor{\bsnm{Sharpe}, \binits{W.F.}}:
\bbtitle{Investments}.
\bpublisher{Prentice-Hall},
\blocation{Hoboken}
(\byear{1978})
\end{bbook}
\endbibitem

\bibitem[\protect\citeauthoryear{Skorokhod}{2005}]{Skorokhod_2005}
\begin{bbook}
\bauthor{\bsnm{Skorokhod}, \binits{A.V.}}:
\bbtitle{Basic Principles and Applications of Probability Theory}.
\bpublisher{Springer},
\blocation{Heidelberg}
(\byear{2005})
\end{bbook}
\endbibitem

\bibitem[\protect\citeauthoryear{Shirvani et~al.}{2020}]{shirvani_2020}
\begin{barticle}
\bauthor{\bsnm{Shirvani}, \binits{A.}},
\bauthor{\bsnm{Stoyanov}, \binits{S.V.}},
\bauthor{\bsnm{Rachev}, \binits{S.T.}},
\bauthor{\bsnm{Fabozzi}, \binits{F.J.}}:
\batitle{A new set of financial instruments}.
\bjtitle{Frontiers in Applied Mathematics and Statistics}
\bvolume{6},
\bfpage{606812}
(\byear{2020})
\end{barticle}
\endbibitem

\bibitem[\protect\citeauthoryear{Tian}{1993}]{tian_1993}
\begin{botherref}
\oauthor{\bsnm{Tian}, \binits{Y.}}:
A modified lattice approach to option pricing
\textbf{13},
563--577
(1993)
\end{botherref}
\endbibitem

\bibitem[\protect\citeauthoryear{Yuen and Yang}{2010}]{yuen_2010}
\begin{barticle}
\bauthor{\bsnm{Yuen}, \binits{F.L.}},
\bauthor{\bsnm{Yang}, \binits{H.}}:
\batitle{Option pricing with regime switching by trinomial tree method}.
\bjtitle{Journal of Computational and Applied Mathematics}
\bvolume{233}(\bissue{8}),
\bfpage{1821}--\blpage{1833}
(\byear{2010})
\end{barticle}
\endbibitem

\end{thebibliography}

\end{document}